\documentclass[aps,superscriptaddress,nofootinbib,showpacs,notitlepage, twocolumn, prb]{revtex4-1}
\usepackage{amsfonts}
\usepackage{amsmath}
\usepackage{amsthm}
\usepackage{braket}
\usepackage{graphics}
\usepackage{hyperref}
\usepackage[dvipsnames]{xcolor}
\usepackage{amssymb}
\usepackage{dsfont}
\usepackage{verbatim}
\usepackage{amsmath,amscd}
\usepackage[all,cmtip]{xy}
\usepackage{multirow}

\usepackage{etoolbox,tikz, tikz-cd}

\usetikzlibrary{cd}

\usepackage[capitalize]{cleveref}

\Crefname{enumi}{Case}{Cases}
\Crefname{subsection}{Subsection}{Subsections}

\definecolor{darkblue}{RGB}{0,0,127} 
\definecolor{darkgreen}{RGB}{0,130,80}
\definecolor{darkred}{RGB}{150,10,10}
\hypersetup{
	colorlinks,
	linkcolor=darkblue,
	citecolor=darkgreen,
	filecolor=red,
	urlcolor=blue,
	pdftitle={},
	pdfauthor={}
}

\graphicspath{{./Figures/}}

\usepackage{tikz,pgfplots}\pgfplotsset{compat=newest}
\usetikzlibrary{external,calc,decorations.pathreplacing,decorations.markings,decorations.pathmorphing,arrows.meta,shapes.geometric}
\newlength\figureheight
\newlength\figurewidth

\usepackage{graphicx}
\usepackage{tabularx}
\usepackage{booktabs}
\usepackage{array}
\usepackage{amsmath}
\usepackage{amsfonts}
\usepackage{amssymb}
\usepackage{amsthm}
\usepackage{enumitem}
\usepackage{permute}
\usepackage{url}
\usepackage{balance}

\DeclareFontFamily{U}{mathb}{\hyphenchar\font45}
\DeclareFontShape{U}{mathb}{m}{n}{
      <5> <6> <7> <8> <9> <10> gen * mathb
      <10.95> mathb10 <12> <14.4> <17.28> <20.74> <24.88> mathb12
      }{}
\DeclareSymbolFont{mathb}{U}{mathb}{m}{n}
\DeclareMathSymbol{\bigast}{2}{mathb}{"06}

\def\XXint#1#2#3{{\setbox0=\hbox{$#1{#2#3}{\int}$}
     \vcenter{\hbox{$#2#3$}}\kern-.5\wd0}}

\theoremstyle{plain}

\theoremstyle{definition}

\newtheoremstyle{remark}
{}   
{}   
{\normalfont}  
{}       
{\itshape} 
{.}         
{5pt plus 1pt minus 1pt} 
{}          

\theoremstyle{remark}

\newlist{alternative}{enumerate}{4}     
\setlist[alternative,1]{label=\arabic*., ref=\arabic*}
\setlist[alternative,2]{label=(\alph*), ref=\thealternativei.(\alph*)}
\setlist[alternative,3]{label=\roman*., ref=\thealternativei.(\thealternativeii).\roman*}
\setlist[alternative,4]{label=\Alph*., ref=\thealternativei.(\thealternativeii).\thealternativeiii.\Alph*}

\setlist[enumerate,1]{label=\arabic*., ref=\arabic*}
\setlist[enumerate,2]{label=(\alph*), ref=\theenumi.(\alph*)}
\setlist[enumerate,3]{label=\roman*., ref=\theenumi.(\theenumii).\roman*}
\setlist[enumerate,4]{label=\Alph*., ref=\theenumi.(\theenumii).\theenumiii.\Alph*}

\AtBeginEnvironment{tikzcd}{\tikzexternaldisable}
\AtEndEnvironment{tikzcd}{\tikzexternalenable}

\newcommand{\drawgenerator}[8]{%
\xymatrix@!0{%
& #8 \ar@{-}[ld]\ar@{.}[dd] \ar@{-}[rr] & & #7 \ar@{-}[ld]  \\%
#1 \ar@{-}[rr] \ar@{-}[dd] &  & #2 \ar@{-}[dd] &            \\%
& #6 \ar@{.}[ld] &  & #5 \ar@{-}[uu] \ar@{.}[ll]       \\%
#3 \ar@{-}[rr] &  & #4 \ar@{-}[ru]                       %
}%
}

\newcommand{\plaquette}[4]{
\xymatrix@!0{%
#1 \ar@{-}[r] \ar@{-}[d]  & #2 \ar@{-}[d] 
\\
#3 \ar@{-}[r]  & #4
}}

\usepackage{floatrow}
\usepackage[caption=false,label font={bf,normalsize}]{subfig}
\definecolor{jon}{RGB}{255,0,0}

\definecolor{arpit}{RGB}{127,0,0}

\definecolor{dom}{RGB}{10,0,100}

\begin{document}

\title{
Bifurcating subsystem symmetric entanglement renormalization in two dimensions
}

\author{Jonathan Francisco San Miguel}
\affiliation{Department of Physics, Stanford University, Stanford, CA 94305, USA}

\author{Arpit Dua}
\affiliation{Department of Physics, Yale University, New Haven, CT 06520-8120, USA}
\affiliation{Department of Physics and Institute for Quantum Information and Matter, \mbox{California Institute of Technology, Pasadena, California 91125, USA}}

\author{Dominic~J. Williamson}
\affiliation{Department of Physics, Stanford University, Stanford, CA 94305, USA} 

\begin{abstract}
\noindent
We introduce the subsystem symmetry-preserving real-space entanglement renormalization group and apply it to study bifurcating flows generated by linear and fractal subsystem symmetry-protected topological phases in two spatial dimensions.
We classify all bifurcating fixed points that are given by subsystem symmetric cluster states with two qubits per unit cell. 
In particular, we find that the square lattice cluster state is a quotient-bifurcating fixed point, while the cluster states derived from Yoshida's first order fractal spin liquid models are self-bifurcating fixed points. 
We discuss the relevance of bifurcating subsystem symmetry-preserving renormalization group fixed points for the classification and equivalence of subsystem symmetry-protected topological phases. 
\end{abstract}

\maketitle

\section{introduction}
The classification of all phases of matter is central to condensed matter physics. 
Over the past several decades, the exploration of gapped quantum phases of matter at zero temperature has led to new and exotic possibilities. 
This progress has been driven in large part by the discovery of deep connections between these phases and quantum codes and computation~\cite{Zeng2015}.
Topological phases of matter~\cite{doi:10.1142/S0217979290000139} have given rise to the field of topological quantum computation~\cite{qdouble,shor1996fault,nayak2008non}, while symmetry-protected topological (SPT) phases of matter~\cite{chen2013symmetry} were found to underlie measurement based quantum computation (MBQC) in quantum wires~\cite{Else2012,Else2012a,Stephen2017,Raussendorf2017}. 
The connection 
has proven reciprocal, as unconventional topological quantum codes~\cite{chamon2005quantum,bravyi2011topological,haah2011local,bravyi2013quantum,vijay2016fracton,Brown2019} have led to the fascinatingly unexpected fracton phases of matter~\cite{Nandkishore2019,Pretko2020}, while universal MBQC resource states~\cite{Raussendorf2001,Briegel2009} have motivated the study of related subsystem symmetry-protected topological (SSPT) phases~\cite{raussendorf2018computationally,you2018subsystem,devakul2018fractal,devakul2018universal,Stephen2018computationally,Williamson2018,Daniel2019,Devakul2020b}.

A leading approach to classifying quantum phases is via stable fixed points of the entanglement renormalization group (ERG)~\cite{Vidal2007}, a carefully controlled form of blocking real-space renormalization designed to preserve zero temperature quantum phases~\cite{haah2014bifurcation}. 
Recently, this picture has been challenged by fracton phases that exhibit exotic bifurcating ERG flows~\cite{haah2014bifurcation,shirley2017fracton,Dua2019b}. This has led to a generalization of the usual notion of ERG fixed-point to also allow self-bifurcating and quotient-bifurcating fixed points, which define coarser equivalence classes of fracton phases~\cite{shirley2017fracton,Dua2019b}. 

While there are no intrinsic topological phases in 1D spin systems, there are nontrivial SPT phases that exhibit topological phenomena with respect to operations that commute with a global symmetry~\cite{Chen2011,SchuchGarciaCirac11}. 
These phases are classified by equivalence classes of symmetry-preserving ERG fixed-points~\cite{gu2009tensor,Chen2011}. 
Analogously, while there are no fracton topological phases in 2D spin systems~\cite{Aasen2020,Haah2020}, there are nontrivial SSPT phases that exhibit similar topological phenomena with respect to subsystem symmetric operators~\cite{you2018subsystem,devakul2018fractal}. To date, the classification of these phases has garnered significant interest and ample progress~\cite{subsystemphaserel,Devakul2018,Shirley2019d,Devakul2020}, but a consensus has not been reached. 
This raises a natural question: what is the nature of the flows generated by SSPTs under subsystem symmetry-preserving ERG?

In this work, we formulate and study the subsystem symmetry-preserving entanglement renormalization group (SSPERG) for both linear and fractal subsystem symmetries. 
We uncover symmetric gapped self-bifurcating and quotient-bifurcating fixed points directly in 2D, reminiscent of the behaviour exhibited by fracton models in 3D~\cite{haah2014bifurcation,shirley2017fracton,Dua2019b}. 
To the best of our knowledge, nontrivial symmetric gapped bifurcating ERG fixed points in less than 3D have not appeared previously in the literature. 

To find bifurcating fixed point solutions we utilize \textit{twist phases}, defined in Refs.~\onlinecite{you2018subsystem,Devakul2018,devakul2018fractal}, as invariants for the SSPERG flows. These invariants provide key constraints on the possible SSPERG flows that allow us to rule out certain a priori possibilities, and in many cases, to find particular bifurcating fixed point solutions. 
Following this approach, we first show that the square lattice cluster state with linear subsystem symmetries is a quotient-bifurcating fixed point under the SSPERG. 
Second, we classify bifurcating fixed points under fractal SSPERG that take the form of first order fractal SPT cluster states with two qubits per unit cell. These include the 2D fractal SPTs defined by Yoshida's first order fractal spin liquids~\cite{yoshida2013exotic} (FSL) which we find to be self-bifurcating.

The paper is laid out as follows. In Section~\ref{sec:background}, we review necessary background topics: the entanglement renormalization group (both in the absence and presence of global symmetry), subsystem symmetries and subsystem symmetry protected topological phases. Next, in Section \ref{sec:sserg}, we combine these topics and describe the subsystem symmetry-preserving ERG. 
In Section~\ref{sec:Examples}, we construct bifurcating SSPERG flows for two models: the square lattice cluster state, and a family of fractal SPT cluster states. 
Finally, in Section~\ref{sec:Discussion}, we conclude and discuss further research directions.

In the appendices, we provide more specific details about the ERG flows in our examples. In Appendix~\ref{app:polynomial}, we briefly review Haah's polynomial notation, which is useful for discussing the ERG. In Appendix~\ref{linearappendix}, we describe the square lattice cluster state, and finally, in Appendix~\ref{appendixfractal}, we discuss fractal SSPTs. All of these examples are also included in the Mathematica notebook in the supplementary material.

\section{Background}
\label{sec:background}
In this section, we first review the entanglement renormalization group~\cite{Vidal2007}, with and without symmetry~\cite{Singh2013,Huang2013}. We then review subsystem symmetries and subsystem symmetry protected topological phases~\cite{you2018subsystem,devakul2018fractal}. Next, we introduce cluster states~\cite{briegel2001persistent,Raussendorf2003}, which provide examples of SSPT phases~\cite{raussendorf2018computationally,you2018subsystem,devakul2018fractal,devakul2018universal,Stephen2018computationally,Williamson2018,Daniel2019}. Finally, we review the twist phase, a topological invariant that can be used to classify SSPT phases~\cite{Devakul2018}.

\subsection{Entanglement renormalization group}
A powerful way to understand critical behavior and define phases of matter is through the renormalization group~\cite{Kadanoff:1966wm,Wilson1975}, which allows for the study of a system at larger and larger scales, removing any physics that is only present at short range. The entanglement renormalization group (ERG) is a class of  renormalization group transformations for systems on a lattice, which coarse-grain the lattice and remove short-range entanglement~\cite{Vidal2007}. A wide range of possibilities exist for ERG fixed points. Conventional fixed points have been described that show self-similarity at increasing scales~\cite{Gu2008,Aguado2008,Konig2009,chen2010local}, but more exotic possibilities, such as \textit{self-bifurcation}~\cite{Evenbly2014c,Evenbly2014,Evenbly2014a} and \textit{quotient-bifurcation}~\cite{Dua2019b}, have also been observed. For example, in phases that self-bifurcate, such as those with fracton topological order~\cite{haah2014bifurcation,shirley2017fracton,Dua2019b}, the system splits into two or more copies of itself after each ERG transformation.

The ERG has been utilized for both topologically ordered phases~\cite{Gu2008,Aguado2008,Konig2009,chen2010local} and symmetry protected phases~\cite{gu2009tensor,Chen2011,Singh2013,Huang2013}. For the latter, the class of RG transformations is restricted to preserve a given symmetry group~\cite{gu2009tensor,Singh2013,Huang2013}. This restriction creates a subdivision of the topologically ordered phase equivalence relation~\cite{gu2009tensor,Chen2011,SchuchGarciaCirac11}. In this section, we describe both variants of the ERG, first considering its application to topological phases without symmetry before moving on to the symmetry-preserving case.

\subsubsection{Entanglement renormalization for topological phases}
\label{sec:ERG}

\begin{figure}[t!]
    \centering
    \includegraphics[width=\columnwidth]{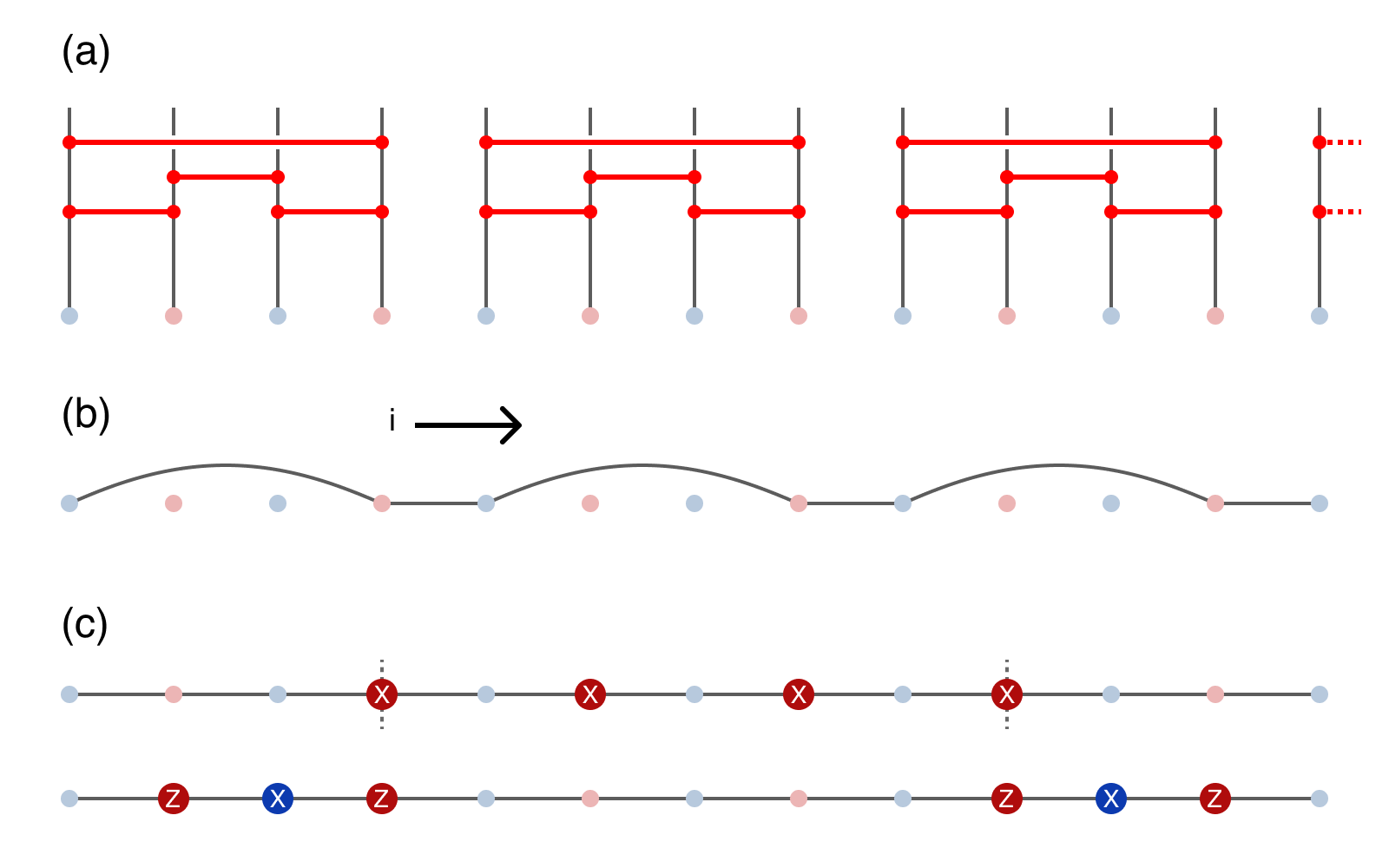}
    \caption{(a) Symmetric LUC taking $H(a)$ (Eq. \ref{1dhamiltonian}) to $H(2a)$ (Eq. \ref{1dcoarsegrain}). Each red line is a $CZ$ gate. (b) Coarse-grained 1D cluster Hamiltonian $H(2a)$. The lines between qubits indicate there is a local term containing both qubits. (c) Calculating twist phases in the 1D linear cluster state, with truncated symmetry $S_1$ (top), and local operators that anticommute with it (bottom).}
    \label{1dclusterfig}
\end{figure}

We consider spin systems defined by local Hamiltonians with a finite energy gap on a regular spatial lattice~$\Lambda$ as follows,
\begin{align}
    H = \sum_{i\in\Lambda} h_i \, ,
\end{align}
where $h_i$ are supported on a local neighborhood contained within a constant distance of site $i$. 
We also consider local unitary circuits (LUC)~\cite{chen2010local}, defined by a series of unitaries $U=U_N \dots U_2 U_1$, where 
\begin{equation}\label{localunitarydef}
    U_n =\prod_{i}\mathcal{O}_{n,i},
\end{equation}
for unitaries $\mathcal{O}_{n,i}$ that have local support, disjoint from one another. The support of the $\mathcal{O}_{n,i}$ and the depth of the circuit $N$ are independent of system size. The existence of a local unitary circuit between two gapped, local Hamiltonians implies a gapped adiabatic path between them, and therefore, that they are in the same quantum phase of matter~\cite{chen2010local}.

The entanglement renormalization group~\cite{Vidal2007}, is a particular class of real-space renormalization transformations generated by coarse-graining operations that enlarge the scale of the lattice, local unitary circuits, and the decoupling of degrees of freedom that are in a trivial tensor product state~\cite{haah2014bifurcation}. 
These operations are defined so as to remain within a given quantum phase of matter. Here our focus is on translation invariant gapped commuting projector Hamiltonians, and so the local unitary circuits are also taken to be translation invariant. By coarse-graining the lattice, we are picking out a subgroup of translations that need to be preserved, and hence allowing for a larger set of local unitaries. 

To describe a general transformation step of the ERG, we start with a local, gapped Hamiltonian $H(a)$ defined on a lattice with lattice constant $a$. 
An ERG transformation can be decomposed into three steps~\cite{Dua2019b}:
\begin{enumerate}
    \item Coarse-grain the lattice; in other words, redefine the unit cell to be size $ca$, for $c$ an integer.
    \item Apply local unitary circuits to remove short-range entanglement.
    \item Project out any trivial and disentangled degrees of freedom.
\end{enumerate}

After performing these steps, what is left is a set of one or more disjoint local, gapped Hamiltonians defined on the coarse-grained lattice. This may be written as
\begin{equation}
H(a) \simeq H_1(ca) + H_2(ca) + ...
\end{equation}
where $\simeq$ stands for quantum phase equivalence, which includes LUCs, decoupling trivial degrees of freedom, and redefinitions of the local Hamiltonian terms that preserve the ground space and the energy gap. 

This process may then be repeated on each of the Hamiltonians $H_i(ca)$. By definition, the ERG preserves quantum phases of matter, and can be used to demonstrate the equivalence of phases on a lattice by coarse-graining at larger and larger scales.

An \textit{ERG fixed point} occurs when the ERG repeats after a constant number of steps. By repeating, we mean that the result of each ERG step can be written using copies of the Hamiltonians from the previous step, but defined at larger and larger lattice scales. 
In this work, we focus on ERG fixed points, or models that run to such fixed points in a finite number of steps. It is worth noting that the ERG may also be defined for generic models that only asymptotically reach their fixed point, but
this is beyond the scope of our current work, and likely requires the application of numerical techniques~\cite{Evenbly2009a}.

The simplest type of fixed point is that of a single Hamiltonian \textit{conventional fixed point}, 
\begin{equation}
    H(a) \simeq H(ca) \, ,
\end{equation} 
relevant for the case of conventional topological phases~\cite{Kitaev2001,Levin2005,Gu2008,Aguado2008,Konig2009}.
Another important possibility is \textit{self-bifurcation}, where a Hamiltonian splits into multiple copies of itself such as 
\begin{equation}
    H(a) \simeq H_1(ca)+H_2(ca) \, ,
\end{equation}
with $H_{1}(ca) \simeq H_{2}(ca) \simeq H(ca)$. 
More generally it is possible to have \textit{quotient-bifurcation} where, instead, $H(ca)\simeq H_1(ca)$, $H_1(ca) \not\simeq H_{2}(ca)$ and $H_{2}(ca)$ self-bifurcates.
These latter possibilities are relevant for fracton topological phases~\cite{haah2014bifurcation,shirley2017fracton,Dua2019b}. 

\subsubsection{Symmetry-Preserving ERG}

In the presence of symmetries, a restricted form of ERG transformations that are symmetry-preserving have previously been considered~\cite{Huang2013,Singh2013}. 
Symmetry-enriched topological phases (SETs)~\cite{barkeshli2014symmetry} are defined by symmetric Hamiltonians under an equivalence relation that allows any gapped adiabatic path \textit{that respects a given symmetry group} $G$, with representation $S(g)$ for $g\in G$.
The well known subclass of symmetry-protected topological phases (SPTs) are defined by further restricting to SETs that are equivalent to the topologically trivial phase when all symmetry conditions are dropped~\cite{Chen2011,SchuchGarciaCirac11}. 
The symmetric phase equivalence condition is ensured if we only consider symmetric Hamiltonians and modify definition~\eqref{localunitarydef} to require that each $\mathcal{O}_{n,i}$ commutes with all of the symmetry operators $S(g_j)$, for all $g_j$ in $G$. In other words, every local unitary operator in the LUC must respect the symmetries~\cite{chen2010local}. It follows that any trivial states which are decoupled during the ERG must be symmetric trivial states.

A simple example of a symmetric ERG fixed point is given by the following Hamiltonian (known as the 1D cluster state~\cite{briegel2001persistent}, (see Section \ref{sec:ssptexample} for more on cluster states):
\begin{equation}\label{1dhamiltonian}
    H(a)=\sum_i Z_{i-1}X_iZ_{i+1}.
\end{equation}
There are two symmetry generators, $S_1$ and $S_2$, consisting of Pauli $X$ operators on even and odd sites, respectively. There exists a local circuit made up of controlled-Z (CZ) gates on sets of four adjacent qubits (see Fig.~\ref{1dclusterfig}), taking this Hamiltonian to
\begin{equation}\label{1dcoarsegrain}
    H(2a) = \sum_i Z_{4i-3}X_{4i}Z_{4i+1}+Z_{4i}X_{4i+1}Z_{4i+4}.
\end{equation}
We refer to the second model as $H(2a)$ because it is essentially the same as $H(a)$, except that it is only supported on half of the qubits (see Fig.~\ref{1dclusterfig}). Its unit cell is also four qubits, instead of two. 
It is easily verified that the local gates given by a product of four CZ gates respect both symmetry generators. Therefore, we have $H(a) \simeq H(2a)$, which recovers the well-known result that the cluster state is a symmetric ERG fixed-point.

A final point about this example is that we may turn the Hamiltonian in Eq. \ref{1dhamiltonian} into a trivial, noninteracting model by adding CZ gates between every pair of adjacent qubits. This does not, however, mean that the original model is in a trivial SPT phase; in fact, it is nontrivial. While such a circuit does \textit{globally} commute with both symmetries, it cannot be decomposed into local operators $\mathcal{O}_{n,i}$ that commute with the symmetries individually. As this example shows, nontrivial SPTs may be trivialized with a globally symmetric circuit if we drop the local symmetry-preserving condition. More generally, the classification of SETs represents a subdivision of the classification of topological states without symmetries.

\subsection{Subsystem symmetry}
The $\mathbb{Z}_2\times \mathbb{Z}_2$ symmetry group of the Hamiltonian in Eq.~\eqref{1dhamiltonian} is an example of a \textit{global symmetry}: the symmetry group does not change with system size. A more exotic possibility is that the symmetry group is generated by subextensive operators, and its cardinality grows with the system size. In this case, it is called a subsystem symmetry~\cite{you2018subsystem,devakul2018fractal}.

Several classifications of subsystem symmetry protected phases have been proposed~\cite{subsystemphaserel,Devakul2018,Shirley2019d,Devakul2020}. Some of these classifications have been based on the twist phase, a topological invariant related to the projective representations of the symmetry group. 
We introduce the twist phase below, alongside the definition of subsystem symmetries. 

\subsubsection{Subsystem symmetry groups}
\label{sec:subsyms}

To give concrete examples of subsystem symmetry, we follow the treatment of Ref.~\onlinecite{Williamson2020a}, and restrict to the case of \textit{locally specified} symmetry operators that consist of Pauli $X$ operators. This class of symmetries is defined as follows: Let $P_Z$ and $P_X$ be the subgroups of the Pauli group generated by all Pauli $Z$ and $X$ operators, respectively. First, write down a subgroup $\Gamma$ of $P_Z$, consisting of local operators. This subgroup is called the \textit{constraints}. Then, take the symmetry group $S$ to be all products of Pauli $X$ operators that commute with the constraints, or
\begin{equation}
    S = \{ s \in P_X \:|\: s\sigma s = \sigma\: \forall \: \sigma \in \Gamma \}.
\end{equation}
This generalizes the notion of a global spin flip symmetry, which can be described in this formalism by choosing constraints that are generated by nearest neighbor two-body Pauli $Z$ operators on a lattice. For example, on the square lattice, these two body operators are  
\begin{align}\label{globalconstraints}
\Gamma=\left\{ Z_{i,j}Z_{i+1,j},Z_{i,j}Z_{i,j+1}, \; \forall i,j\right\} ,
\end{align}
where $Z_{i,j}$ acts as Pauli $Z$ on site $x=i, y=j$.

Let $S_{loc}$ be the subgroup of $S$ generated by all local operators in $S$. Next, consider the group $S_{nl} = S / S_{loc}$. The representatives of nonidentity elements of this group form a minimal set of nonlocal elements of $S$, modulo local symmetries. 
Since this group is Abelian we can decompose it as $S / S_{loc}\cong \mathbb{Z}_2^{K}$ by choosing a set of generating elements. For the discussion below, we assume we have chosen a set of generators of $S_{nl}$ with minimal support. 

If the support of a nonlocal symmetry generator grows extensively (at a rate linear in the system size), it is a global symmetry. If it grows at a rate slower than linear, it is a subsystem symmetry. If the system is translationally invariant, the number of subsystem symmetry generators in $S_{nl}$ grows subextensively with the system size.

We now consider a few examples of subsystem symmetry, by specifying their local constraints $\Gamma$. In all cases, we assume a square lattice. The first example is linear symmetry. Consider the constraints
\begin{align}\label{linearconstraints}
\Gamma=\left\{ Z_{i-1,j}Z_{i+1,j}Z_{i,j-1}Z_{i,j+1}, \; \forall i,j\right\}.
\end{align}
The symmetries that commute with these constraints are Pauli $X$ operators acting on diagonal lines, $x\pm y = c$, through the lattice.  These symmetries are nonlocal and scale subextensively in the system size, and are therefore subsystem symmetries.

\begin{figure}
    \centering
    \includegraphics[width=\columnwidth]{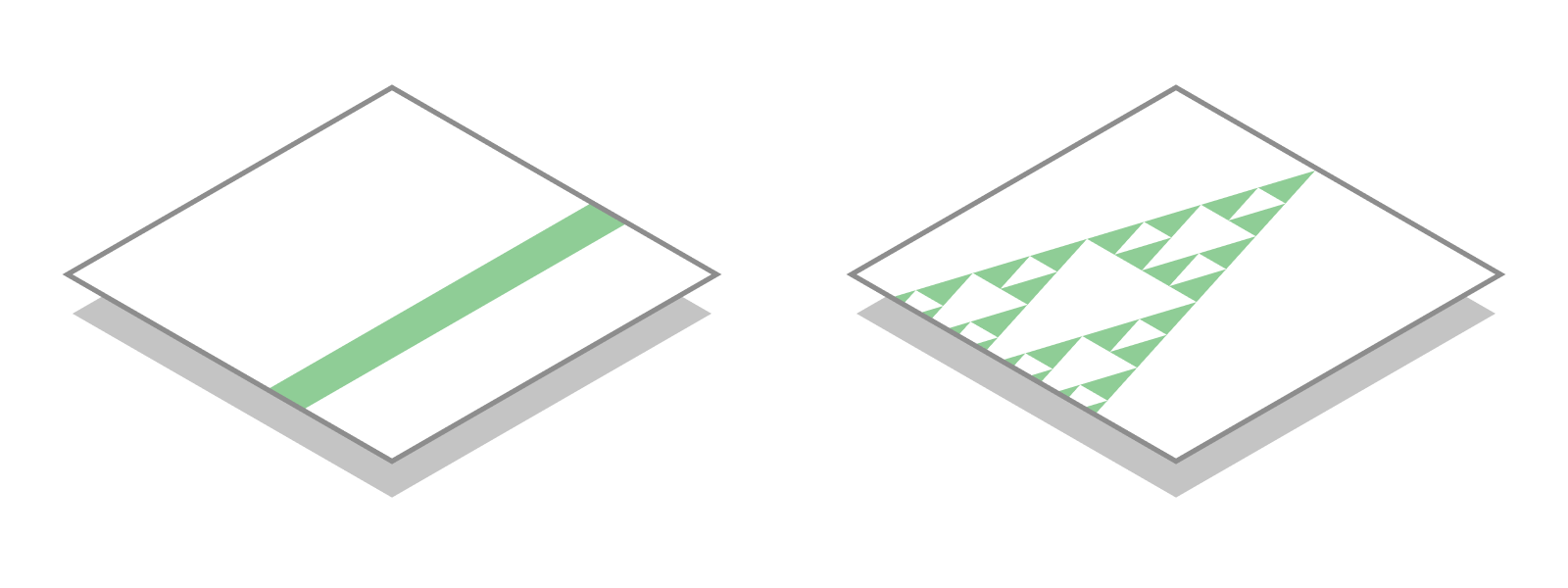}
    \caption{Schematic pictures of subsystem symmetry. Linear (left) and fractal (right) symmetries on a square system.}
    \label{fig:subsystemex}
\end{figure}

To count the number of independent symmetries, consider a diamond of side length $L$ (see Fig.~\ref{fig:subsystemex}). Then there are $2L-1$ independent symmetries, since the product of all symmetries in one direction is equal to the product of all symmetries in the perpendicular direction. The number of symmetries therefore scales as the square root of the number of qubits, again subextensive. This square root scaling is a general property of translationally invariant two dimensional subsystem symmetries\cite{you2018subsystem}.

A more exotic possibility is fractal subsystem symmetry~\cite{PhysRevB,devakul2018fractal}. Consider the constraints 
\begin{align}\label{fractalconstraints}
    \Gamma=&\left\{ Z^{(A)}_{i,j}Z^{(A)}_{i,j+1}Z^{(A)}_{i+1,j+1}, \; \right.
    \nonumber 
    \\
     &\qquad \left. Z^{(B)}_{i,j}Z^{(B)}_{i,j-1}Z^{(B)}_{i-1,j-1}, \; \forall i,j\right\},
\end{align}
where $A$ and $B$ are two sublattices. The symmetries resulting from these constraints have the form of  discrete Sierpinski fractals (see Fig.~\ref{fig:subsystemex}). These are again subsystem symmetries. If we put a Sierpinski fractal on a square lattice of side length $2^l$, the support of the fractal scales as $N^{\log(3)/2}$, where $N$ is the number of lattice sites. This is one example of a larger class of fractal symmetries, which we explore in more depth in Section~\ref{sec:fractal}.

Both Eq. \ref{linearconstraints} and Eq. \ref{fractalconstraints} are examples of a more general type of constraint, defined by a bipartite graph $\textbf{G}=(V,E)$. For such a graph, we may define
\begin{align}\label{graphconstraints}
\Gamma = \left\{\prod_{v' \in N(v)}Z(v') \,, \; \forall \; v \in V\right\},
\end{align}
where $N(v)$ are the neighbors of $v$ according to the edges of $\textbf{G}$. To reproduce the constraints in Eq.~\eqref{linearconstraints}, we simply use the nearest neighbor graph of a square lattice. For Eq.~\eqref{fractalconstraints}, on the other hand, we use a honeycomb lattice.

While Eq. \ref{graphconstraints} does not require a bipartite graph to be defined, we generally restrict to this case. The reason is that on a bipartite graph, the constraints may be split into two sets, each acting entirely on one of the two vertex partitions. Therefore, the symmetries may also be split into two disjoint sets as well. This property is useful for constructing symmetric local unitary circuits (see Section \ref{sec:cluster} below).

\subsubsection{Examples of SSPTs}
\label{sec:ssptexample}
We can generate a short-range entangled Hamiltonian with the symmetry group specified by Eq. \ref{graphconstraints}, for any graph $\textbf{G} = (V,E)$, using a class of models called cluster states~\cite{briegel2001persistent,Raussendorf2003}. The Hamiltonian is given in terms of this graph by
\begin{equation}\label{generalcluster}
    H=\sum_{v\in V}X_v\prod_{v'\in N(v)}Z_{v'},
\end{equation}
We remark that the Hamiltonian in  Eq.~\eqref{1dhamiltonian} is a simple example of a cluster state. In that case, the graph $\textbf{G}$ is just a 1D chain.

To see why the symmetry group of the cluster state Hamiltonian is given by Eq. \ref{graphconstraints}, note that the Pauli $Z$ part of the local operators in Eq. \ref{generalcluster} are the nearest neighbors of each vertex, or exactly the generators of Eq. \ref{graphconstraints}. To see why it is short-range entangled, we construct the ground state:
\begin{equation}\label{clustergroundstate}
    |0\rangle_{clus} = \prod_{e\in E} CZ(e) |+\rangle^{\otimes N},
\end{equation}
Where $CZ(e)$ denotes a $CZ$ operator applied to both of the qubits sharing edge $e$, and $|+\rangle = \frac{1}{\sqrt{2}}\left(|0\rangle + |1\rangle\right)$. By applying a $CZ$ gate to each edge (as with the Hamiltonian in Eq. \ref{1dhamiltonian}), this ground state simply becomes the product state $|+\rangle^{\otimes N}$. Therefore, as long as the edges of the graph have a finite maximum length with respect to the spatial lattice, the Hamiltonian of Eq. \ref{clustergroundstate} is short-range entangled.

As an example of a cluster state with subsystem symmetry, consider the nearest neighbor graph on the square lattice. The corresponding cluster state is
\begin{equation} \label{2dcluster1}
    H = \sum_{ij} X_{ij}Z_{i+1,j}Z_{i-1,j}Z_{i,j+1}Z_{i,j-1} .
\end{equation}
Since this cluster state is defined on a square lattice, the symmetries are precisely those given by Eq.~\eqref{linearconstraints}. Using the twist phase (introduced below), once can verify that this is indeed a nontrivial SSPT phase~\cite{raussendorf2018computationally,you2018subsystem}. The Sierpinski fractal symmetries of the last section (Eq. \ref{fractalconstraints}) may also be realized on a cluster state, but this time using a honeycomb lattice~\cite{devakul2018fractal,devakul2018universal} for the graph $\textbf{G}$.

By utilizing cluster states, we can construct Hamiltonians with subsystem symmetries. However, we have not yet discussed if, and why, these Hamiltonians lie in nontrivial symmetry protected phases, for which there exists no smooth local symmetry-preserving adiabatic path to a product state. To do so, we first need to introduce topological invariants of SSPTs, such as the twist phase considered below.

\subsubsection{Twist phases and classification of SSPTs}
\label{sec:twistphase}

The twist phase is a function that takes two symmetry group elements and returns a $U(1)$ phase. It forms a topological invariant that can be used to distinguish between trivial and nontrivial SSPT phases, or between different nontrivial phases. We remark that the twist phase was originally introduced for global symmetries~\cite{you2018subsystem,devakul2018fractal}.

To calculate twist phases, we first take any symmetry operator $S_1$, and truncate it to an operator $S_{1,\geq}$ that only acts nontrivially inside a strip of finite width:
\begin{equation}
    S_{1,\geq}=
    \begin{cases}
    S_1, & x \in [0,x_{cut}]\\
    \mathbb{I}, & \textrm{elsewhere}
    \end{cases},
\end{equation}
where $x_{cut}$ is some finite value. 
Next, due to the locality of $H$, and the local symmetry condition it obeys, 
\begin{equation}
    S_{1,\geq}^\dagger H S_{1,\geq} = V_0^\dagger V_{x_{cut}}^\dagger H V_{x_{cut}}V_0,
\end{equation}
where $V_0$ and $V_{x_{cut}}$ are unitaries that are exponentially localized near $x=0$ and $x=x_{cut}$, respectively (the above equation holds strictly for the cluster state models we consider in this work, while for more general models it only holds in the ground state subspace). 
It can be shown that the operators $V_0$ and $V_{x_{cut}}$ form projective representations of the symmetry group at each edge of the truncated symmetry, $0$ and  $x_{cut}$. Finally, we can define the twist phase $\Omega^{(x)}(S_1,S_2)$ for $S_1$ and $S_2$ by
\begin{equation}\label{twistphasedef}
    S_2V_0 = \Omega^{(x)}(S_1,S_2)V_0S_2.
\end{equation}
We can also define a separate set of twist phases, $\Omega^{(y)}$,  by cutting along the $y$ direction. This definition does not depend~\cite{you2018subsystem,devakul2018fractal} on the placement of the cuts or the choice of localized edge operators $V$. It only depends on the symmetries $S_1$ and $S_2$. The set of possible twist phases for a given symmetry group $G$ form a group under multiplication.

The group of twist phases can be characterized on a 1D or quasi-1D geometry (such as a cylinder). On such a geometry, the symmetry operators $S(g)$ with local support (in the infinite size limit) have trivial twist phases. It is clear that the elements of $G$ corresponding to these symmetries form a normal subgroup, which we call $G_C$.

It can then be shown that the group of twist phases under multiplication is isomorphic to the second cohomology group~\cite{Devakul2018} of the symmetry group modulo local symmetries, $\mathcal{H}^2[G/G_C,U(1)]$. Every Hamiltonian with symmetry group $G$ has twist phases corresponding to an element of this group, which cannot change under symmetric adiabatic transformations~\cite{subsystemphaserel,Stephen2018computationally}. Therefore, if a LUC is phase preserving, it must preserve twist phases for all pairs of symmetries, along both the $x$ and $y$ directions.

We remark that $x$ or $y$ twist phases alone may provide incomplete information about subsystem symmetries. The reason is that subsystem symmetry generators may be ``line-like", or compact in one direction and extended in the other. Such generators only have nontrivial twist phases along one of the directions. An example of this is the 2D linear cluster state, whose twist phases are calculated in Appendix~\ref{linearappendix}.

As a warm-up, we calculate the twist phase associated to the global symmetry group of the Hamiltonian in Eq.~\eqref{1dhamiltonian}. 
We first consider the truncated symmetry operator $S_{1,\geq}$,
\begin{align}
     S_{1,\geq}=\prod_{i=0}^{i_{cut}/2}X_{2i}.
\end{align}
Without loss of generality, we may assume $i_{cut}$ is even. This operator only anticommutes with two local operators in the Hamiltonian in Eq. \ref{1dhamiltonian}: $Z_{-2}X_{-1}Z_0$ and $Z_{i_{cut}}X_{i_{cut}+1}Z_{i_{cut}+2}$ (Fig. \ref{1dclusterfig}). Another operator which anticommutes with the same local operators is $Z_{-1}Z_{i_{cut}+1}$. Since this operator is the product of two local operators at each cut, we may write
\begin{equation}
    V_0=Z_{-1}, \quad V_{i_{cut}}=Z_{i_{cut}+1}.
\end{equation}
Each of these edge operators anticommutes with $S_2$, and therefore $\Omega(S_1,S_2)=-1$. This shows that the Hamiltonian in Eq.~\eqref{1dhamiltonian} is a nontrivial SPT under the global symmetry. 
Calculations for SSPT twist phases in 2D are more complicated, and we have included detailed examples in Appendices~\ref{linearappendix} and~\ref{appendixfractal} for the interested reader. 
There, we show that the examples from the previous section have nonzero twist phases, and are therefore nontrivial SSPTs.

Twist phases have been shown to completely classify SPTs in one spatial dimension~\cite{Chen2011,SchuchGarciaCirac11}. To what extent this remains true in higher dimensions, however, is an open question. Even in light of this limitation, twist phases remain a useful tool for the classification of SSPTs as they can be used to show that two models are in inequivalent SSPT phases. 

As a final point, all of the nontrivial SSPT models we have discussed so far are, by their nature, trivial topological phases when no symmetry is enforced. This follows from the fact that a cluster state may be trivialized by placing a CZ on every edge of the graph, as discussed in Section \ref{sec:ERG}. 
In two dimensions, we expect that all nontrivial bifurcating symmetric ERG flows in gapped phases originate due to SSPTs. 
This is because in 2D only conventional topological phases, which contain unique fixed points, are possible~\cite{Aasen2020,Haah2020}, and there are no nontrivial SSET phases beyond stacking an SSPT with a decoupled topological order~\cite{Stephen2020}. 
We remark that in three or higher dimensions, it is possible to have nontrivial gapped fracton topological phases that contain bifurcating ERG fixed points~\cite{haah2014bifurcation,shirley2017fracton,Dua2019b}. It is also possible to enrich conventional topological phases to form nontrivial SSET phases~\cite{Stephen2020}. 
While we have focused on the simplest nontrivial setting of two dimensional phases in this work, the bifurcating ERG of subsystem symmetry-enriched phases presents an interesting avenue for future study.

\section{Subsystem symmetry-preserving ERG}
\label{sec:sserg}
In this section, we mention new possibilities that arise when we restrict the ERG to preserve subsystem symmetries, and extend the definitions given in Section~\ref{sec:ERG} to account for these possibilities. The key difference is that the subsystem symmetry-preserving ERG must account for both the flow of the Hamiltonians as well as that of the symmetries. We also discuss specific procedures for finding symmetric local unitaries for cluster states on bipartite graphs, a family that includes all of the examples considered in this work.

\subsection{Characterizing subsystem symmetry-preserving entanglement renormalization}
\label{sec:ssergdef}

To define the subsystem symmetry-preserving ERG, as before, we only allow symmetry-preserving local gates in the decomposition of a local unitary circuit introduced in Eq.~\eqref{localunitarydef}. This requirement does not depend on whether the symmetry group has subsystem or global symmetries. The main difference is that now, unlike in the case of global symmetry, we must consider nontrivial flows for both the Hamiltonian and the symmetry group.

To see what is meant by a flow of the symmetry group, and why this is different in the case of subsystem symmetry, we first discuss what happens to general symmetries under coarse-graining. Consider a Hamiltonian $H(a)$ that is symmetric under the group $G(a)$, and a symmetry-preserving LUC that takes $H(a)$ to $H_1(ca)+H_2(ca)+...$ Each of these coarse-grained Hamiltonians $H_i(ca)$ is supported on a disjoint set of qubits $Q_i$, and is symmetric under $G$\footnote{It is also possible to have $S^\dagger H_i S = -H_i$ for symmetry operators $S$ in the Pauli group. However, all of the examples in this work bifurcate into at least one copy of the original model, which excludes this possibility.}.

We define $G_i(ca)$ to be the restriction of the symmetry group to the subset of qubits $Q_i$. To be precise, we first assume on-site symmetry operators, which act as tensor products of single-site operators (this holds for the cluster state examples considered in this work). We write the action of the operator $S$ on qubit $a$ as $S|_a$. The action of $G_i(ca)$ is given by the operators
\begin{equation}\label{coarsegrainhom}
    S_i(g)|_a = \begin{cases}
        S(g)|_a, &a \in Q_i \\
        I, &a \notin Q_i
    \end{cases},
\end{equation}
where $S(g)$ is a symmetry operator and $g$ is any element of $G(a)$. 
For convenience, when defining $G_i(ca)$ we further quotient out any group element that acts trivially on the whole $Q_i$ sublattice, i.e. $S(g)|_a = I,$ for all $a \in Q_i$. 

We now want to define an equivalence relation, $\approx$, between the symmetry groups at different scales, $G_i(ca)$ and $G(a)$, that is consistent with the equivalence relation on Hamiltonians.  
For a self-bifurcating fixed-point our definition  should satisfy $G_i(ca) \approx G(ca)$ for all $i$, for example.

To establish this definition, we first fix a surjective map $\varphi$ from the original lattice to the coarse-grained sublattice $Q_i$ by multiplying each of the original unit vectors by the coarse-graining factor $c$ to obtain a set of coarse-grained unit vectors that generate $Q_i$. This identifies a qubit at site $(x,y)$ in the original coordinates with a qubit at the site $(x,y)$ in the coarse-grained coordinates. If there are multiple qubits per site of the original lattice, we must also specify how $\varphi$ acts on these. For this, we take the mapping from the identification of the Hamiltonians. For example, if $H(a)$ bifurcates into a copy of itself, ${H_i(ca) = H(ca)}$, then we require that $H(a)$ has the same action on qubit $q$ as $H(ca)$ has on $\varphi(q)$. If $H(a)$ bifurcates into some other Hamiltonian $H'(ca)$ (with the same size unit cell), we can perform a similar mapping, by writing $H'(a)$ on the original lattice and identifying it with $H'(ca)$ on the coarse-grained lattice.

We then say that $G_i(ca)$ is equivalent to $G(a)$, denoted $G_i(ca) \approx G(a)$, if there exist two sets of generators $\{g^{(j)}\}\in G(a)$ and $\{g^{(j)}_i\}\in G_i(ca)$, such that
\begin{align}\label{subsystemequiv}
    S(g^{(j)})|_a = S_i(g_i^{(j)})|_{\varphi(a)},
\end{align}
for all $j$ and $a$. We remark that this relation is stronger than a simple isomorphism of groups. It is a restriction on not only the group structure, but the representations of the group as well. This restrictive definition rules out general local unitaries, which are not symmetric, but still preserve the group structure of the symmetries.

As an example, we consider this equivalence relation for the 1D cluster state from Eq.~\eqref{1dhamiltonian}. We first define $\varphi$ based on the identification of $H(2a)$. In this case, the identification from coarse-graining by a factor of two is
\begin{align}
    \varphi(i) = \begin{cases}
    2i, &i \textrm{ even} \\
    2i + 1, &i \textrm{ odd}
    \end{cases}.
\end{align}
Under this map, it is easy to see that the operators $S_1$ and $S_2$ are preserved, and therefore, $G(2a) \approx G(a)$. In general, however, the symmetry group after coarse-graining may not always be equivalent to the original symmetry group. 

The main difference between subsystem and global symmetries appears at fixed points. In either case, coarse-graining defines a homomorphism from $G(a)$ to $G_i(ca)$, specified by Eq. \ref{coarsegrainhom}. For global symmetries, since $|G_i(ca)| \leq |G(a)|$, and $|G(a)|$ is fixed at all sizes, the only possible fixed point symmetry group is of the form encountered with the Hamiltonian in Eq. \ref{1dhamiltonian}, $G_i(2a) \approx G(a)$. 
For subsystem symmetries, however, not only is it possible to have $G_i(2a) \not\approx G(a)$, but this can repeat indefinitely in the infinite size limit. This is due to the subextensive scaling of the subsystem symmetry group, which allows for homomorphisms with a nontrivial kernel at all lattice scales.

We denote the flow of the symmetry group and the Hamiltonians together using ordered pairs; for example,
\begin{align}
    (H(a), G(a)) \simeq (H_1(ca), G_1(ca)) + (H_2(ca), G_2(ca)) + ...
\end{align}
We now refine our definition of SSPERG fixed points as follows: for the standard ERG we required that once a fixed point has been reached, the Hamiltonian at all subsequent steps can be written as a disjoint set of $H_i$ that have appeared at previous steps with smaller lattice scales. For an SSPERG fixed point, however, we require that every pair $(H_i(ca), G_i(ca))$ has shown up at earlier steps at smaller lattice scales, using the equivalence relation in Eq. \ref{subsystemequiv} for the symmetries. In particular, in a self-bifurcating fixed point, we must have that each $G_i(ca) \approx G(a)$.

This definition is necessitated by the possibility that the symmetry group changes under application of the ERG. If a symmetry group acts differently on two Hamiltonians, the standard definition of SPT phase equivalence cannot be directly applied to compare them. 
For an ERG flow to preserve subsystem symmetric phases, we need a notion of equivalence for both the the Hamiltonian and the symmetry group at each level of coarse-graining. The first is already provided in the conventional ERG; the second is given by definition~\ref{subsystemequiv}. 

So far, we have discussed the possibility of the symmetry group changing under the ERG flow, allowed due to subextensive scaling. Another possible consequence of this scaling is bifurcation of the symmetry group. In particular, is it possible that Eq. \ref{coarsegrainhom} can define more than one homomorphism (for different $i$) with different kernels? In this work, we find that this is possible. Subsystem symmetry groups can bifurcate, and in fact, do so generically. 
This can be seen with a simple scaling argument. 

For this argument, we put our system on a square of side length $L$. The number of qubits then scales as $L^2$, but the number of symmetry generators scales as $L$. If we coarse-grain both dimensions by a factor of two, we expect each $G_i(2a)$ to have half the number of generators. However, to have all of the symmetries act nontrivially, we need at least two of the $H_i(2a)$ to be in nontrivial SPT phases. The homomorphisms from $G(a)$ to each $G_i(2a)$ must have different, nontrivial kernels for this to be true. From this scaling, we expect that bifurcation (though not necessarily self-bifurcation) is generic in SSPTs. While we focus on the simplest nontrivial setting of two dimensions in this work, a similar scaling argument applies in higher dimensions as well.

\subsection{Symmetric local gates for cluster states}
\label{sec:cluster}

In order to perform each step of the ERG, we require symmetric local unitary circuits. The problem of finding symmetric local unitary circuits for an arbitrary symmetry group may be difficult in general. For cluster states defined on a bipartite graph, however, we have found a simple algorithm to generate the relevant circuits.

For this algorithm, consider a cluster state defined on a graph $\textbf{G}=(V,E)$, with vertices partitioned into sets $A$ and $B$. As mentioned in Section~\ref{sec:ssptexample}, the symmetry group is generated by Pauli X operators that act exclusively within either set $A$ or $B$.

\begin{figure}[t!]
    \centering
    \includegraphics[width=\columnwidth]{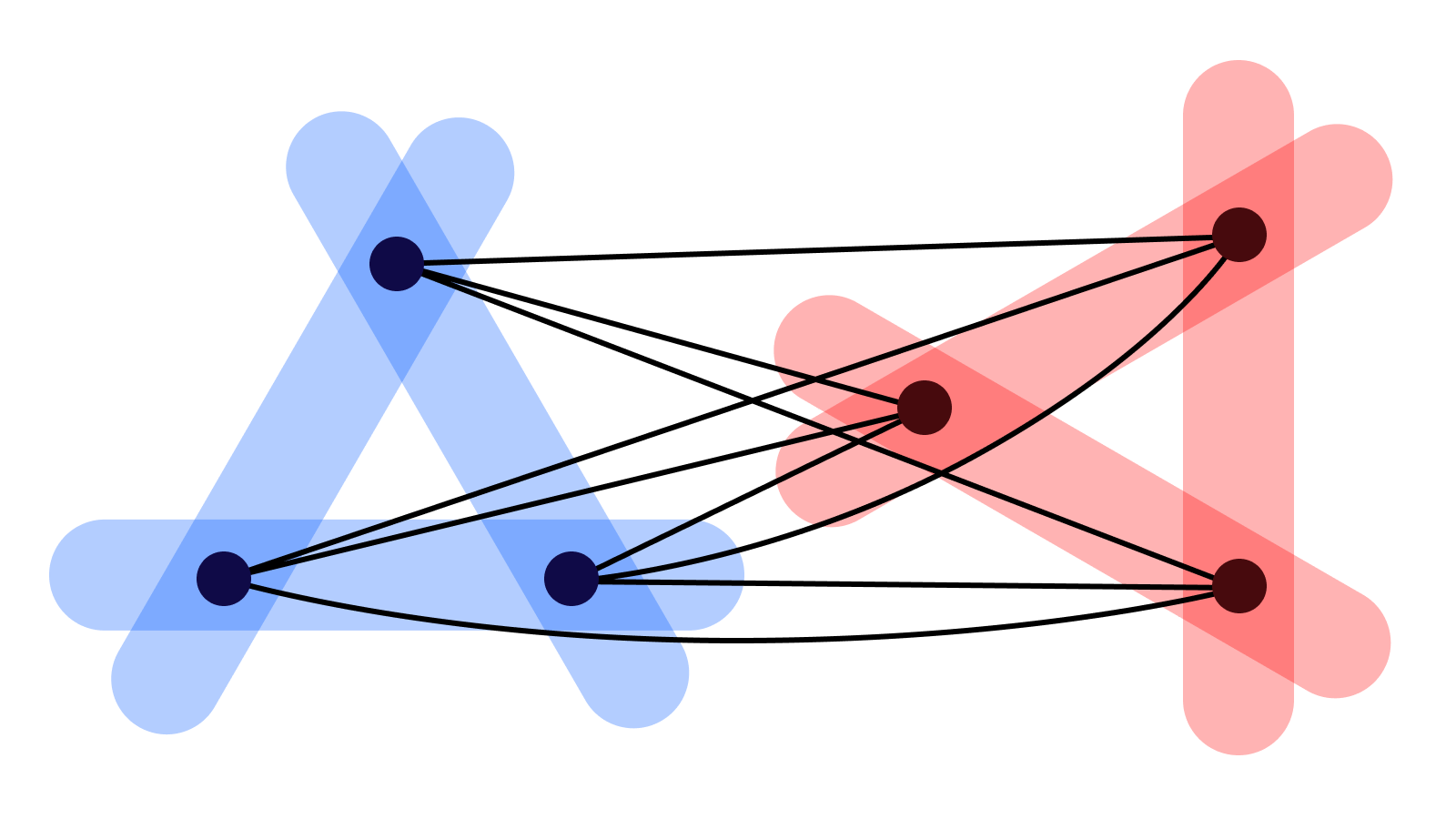}
    \caption{A schematic picture of a local symmetric gate. Blue and red shaded areas represent symmetries on the $A$ and $B$ sublattices, respectively. Dots represent qubits, and lines are CZ gates connecting them.}
    \label{localguatesguidefig}
\end{figure}

The first step is to find two local sets of qubits, $Q_A$ and $Q_B$, lying in $A$ and $B$ respectively, with all symmetries acting on an even number of qubits in each set. Next, add CZ gates from every qubit in $Q_A$ to every qubit in $Q_B$. Such a circuit necessarily preserves the symmetries, and, as long as $Q_A$ and $Q_B$ are local, is local itself. From Eq. \ref{clustergroundstate}, we can see that this takes a cluster state to a new cluster state, defined by
\begin{align}
    \textbf{G} = (V, E + E'),
\end{align}
Where $E'$ is all of the edges between $Q_A$ and $Q_B$, and addition is taken mod 2 over the edges. 
This type of symmetric circuit is depicted schematically in Fig.~\ref{localguatesguidefig}. 
We have, in fact, already given a simple example of such a circuit in Fig.~\ref{1dclusterfig}.

By enumerating all possible $Q_A$ and $Q_B$, we may create an infinite set (in the infinite size limit) of symmetric local unitary circuits. This set becomes finite modulo translations, however, if we only consider unitary circuits with some fixed locality. By fixed locality, we mean restricting the support of $Q_A$ and $Q_B$, as well as the distance between them, to be less than some fixed finite value. We may then search through these unitary circuits to find gates that, when applied in a translationally invariant manner, could implement an ERG flow; for example, by disentangling certain degrees of freedom.

In practice, we find that it is easier to first limit the possible ERG flows using the twist phase. For $c=2$, this limiting procedure works as follows: first, search for models $H_i(2a)$ such that $\sum_i H_i(2a)$ has the same twist phases as $H(a)$.  To simplify this stage, we attempt to find copies of the original Hamiltonian, $H_i(2a) = H(2a)$. These copies of the original Hamiltonian may not be sufficient to satisfy the twist phase constraint. In this case, we are left with additional non-equivalent Hamiltonians, for which we then repeat the same process until a closed ERG is found. We remark that this decomposition may not be unique.

Next, we search for symmetric local circuits that take $H(a)$ to $\sum_i H_i(2a)$. To find these circuits, we first enumerate sets of qubits $Q_A$ and $Q_B$, as defined in Section \ref{sec:cluster}. However, we are aided in this search by the locality of $H_i(2a)$. In all of the examples in this work, we have found that we only need to consider gates with support no larger than the support of the largest local terms in $H_i(2a)$. Such sets are easy to search through by hand.

While the procedure outlined above has two steps, it is worth noting that we have not found a case where simply preserving the twist phases is insufficient. In other words, we have not found $H_i(2a)$ such that $\sum_i H_i(2a)$ has the same twist phase as $H(a)$, but there exists no symmetric LUC taking $H(a)$ to $\sum_i H_i(2a)$. It may be possible that twist phases are complete invariants for SSPT phases, but, as mentioned earlier, this remains unproven beyond one spatial dimension. 

\section{Examples of SSPERG}
\label{sec:Examples}

In this section, we find ERG fixed points for two examples of subsystem SPT phases: the linear cluster state from Eq. \ref{2dcluster1}, and the fractal SPTs (FSPTs), a class of Hamiltonians that realize multiple phases with fractal symmetries. For fractal SPTs in particular, we find fixed points for all first order FSPT cluster states with two qubits per unit cell. 

We first show that the linear cluster state splits into a copy of itself and a second, self-bifurcating model. For the fractal SPTs, meanwhile, we find three possible inequivalent fixed points. Previously, fractal SPTs have been classified by the twist phases \cite{devakul2018fractal}. We show that certain fractal SPT phases may also be classified by their fixed points under the ERG.

\subsection{Linear subsystem symmetry in two dimensions}

We now discuss the 2D linear cluster state from Eq. \ref{2dcluster1}, reproduced below for convenience:
\begin{equation}
    H = \sum_{ij} X_{ij}Z_{i+1,j}Z_{i-1,j}Z_{i,j+1}Z_{i,j-1}.
\end{equation}
This model has been shown to be a nontrivial SSPT\cite{subsystemphaserel}; its twist phases are given in Appendix \ref{linearappendix}.\par

\begin{figure}
    \centering
    \includegraphics[width=\columnwidth]{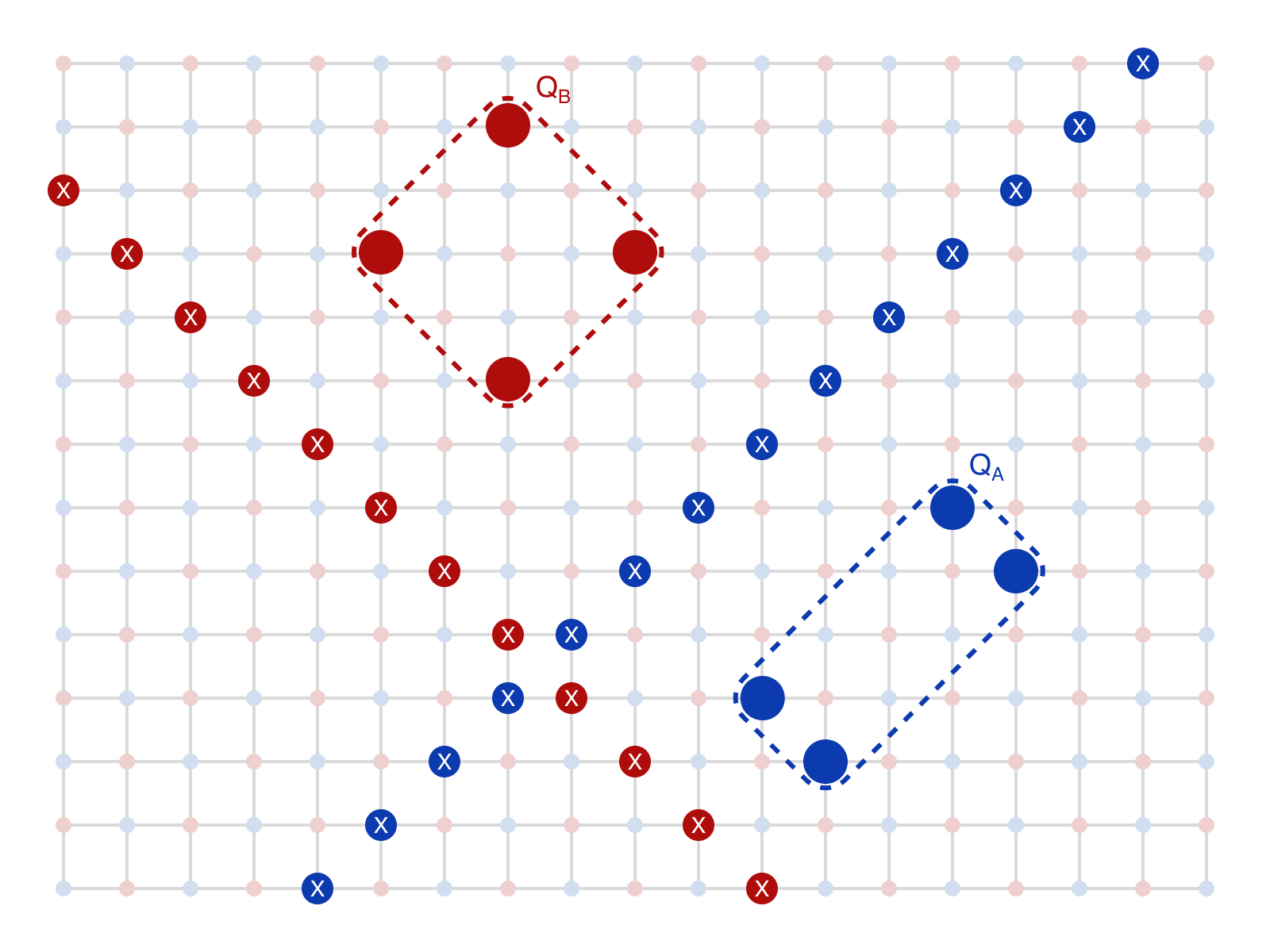}
    \caption{Linear cluster state symmetries (diagonal lines) and examples of sets $Q_A$ and $Q_B$, used to construct symmetric local unitaries.}
    \label{linfig}
\end{figure}

Upon coarse-graining by a factor of two, we find that the 2D linear cluster state splits into a copy of itself and an inequivalent, self-bifurcating model, namely:
\begin{align}\label{linearfixedpoint}
H_{lin}(a) &\simeq H_{lin}(2a) + H_{B}(2a),
\\
H_{B}(a) &\simeq H_B(2a) + H_B(2a).
\end{align}
The definition of the Hamiltonian $H_B$ is provided in Appendix \ref{linearappendix}. A simple set of symmetric gates may be used to generate this ERG. As mentioned in Section \ref{sec:cluster}, we need to find sets of qubits $Q_A$ and $Q_B$; these consist of the vertices of diamonds that have edges on diagonal lines, see Fig.~\ref{linfig}. The exact combination of these gates needed to implement Eq. \ref{linearfixedpoint} is provided in Section 4 of the Mathematica notebook in the supplementary material. 

It is worthwhile considering whether there are other possibilities for the ERG besides Eq.~\eqref{linearfixedpoint}. First, could $H_{lin}(a) \simeq H_{lin}(2a)$, as with the $1D$ case? This turns out to be impossible. In fact, the relation $H_{lin}(a) = H'(2a)$ can never be satisfied for any $H'$ that acts on one qubit per coarse-grained unit cell. It can be shown that every symmetry in the linear cluster state has a nontrivial twist phase with some other symmetry (Appendix \ref{linearappendix}). If we place $H'$ on only one qubit per coarse-grained unit cell (every fourth qubit of the square lattice), however, there are always symmetries that don't act on it at all. These symmetries cannot have a nontrivial twist phase. Therefore, $H_{lin}$ must bifurcate. This argument is simply a more precise restatement of our scaling argument in Section~\ref{sec:ssergdef}.

In Appendix~\ref{linearappendix} we show that the linear cluster state cannot self-bifurcate, which means that any ERG fixed point requires a second inequivalent B model. This is a simple example of a \textit{quotient-bifurcating fixed point}, where a Hamiltonian does not self-bifurcate, but still returns to itself modulo self-bifurcating Hamiltonians under the ERG.

\subsection{Fractal subsystem symmetry in two dimensions}
\label{sec:fractal}
\begin{figure}
    \centering
    \includegraphics[width=\columnwidth]{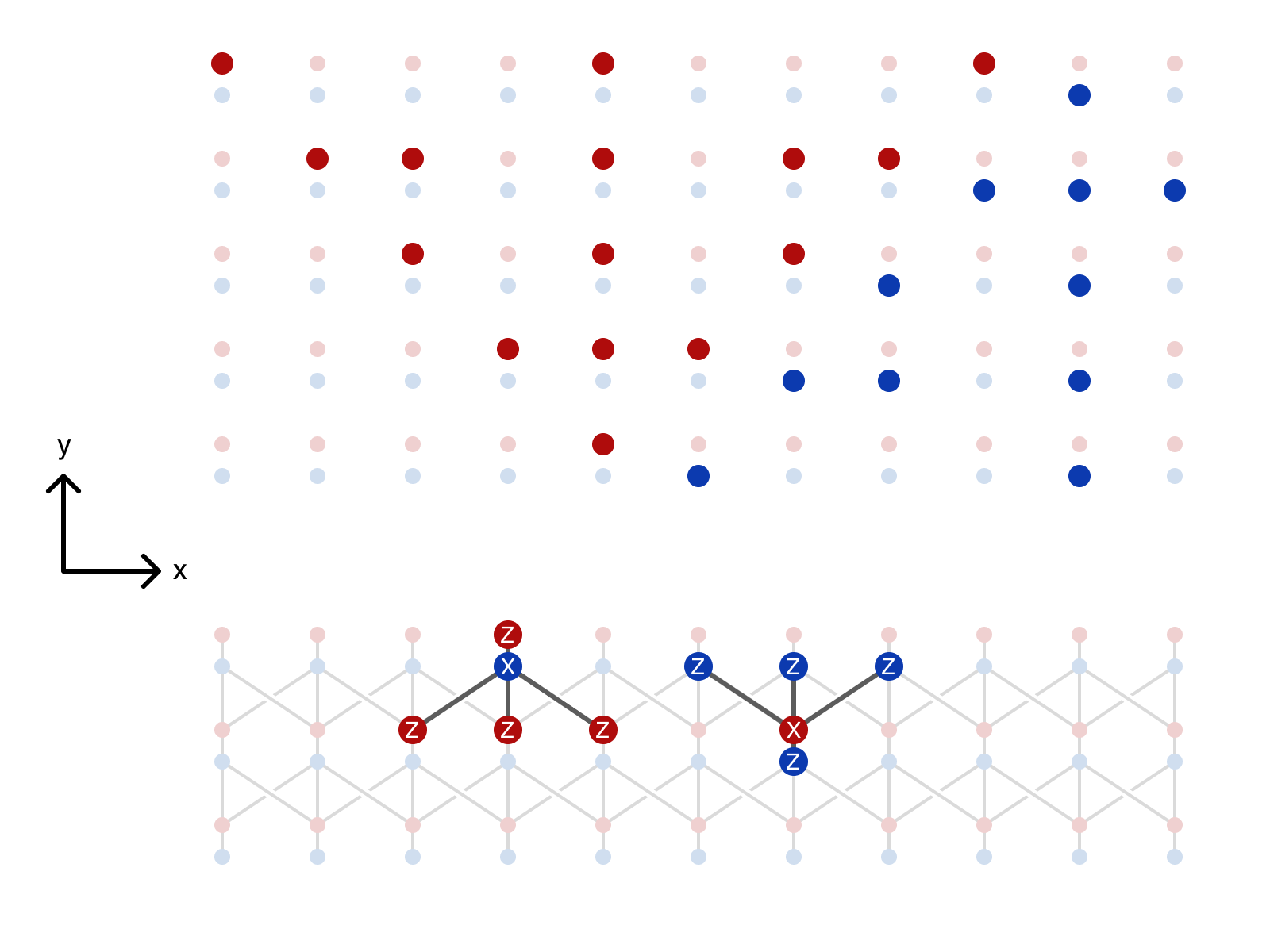}
    \caption{Top: examples of fractal symmetry operators corresponding to the polynomial $f(x)=1+x+1/x$. One example acts as $X$ on the $A$ sublattice (blue), and  the other as $X$ on $B$ (red) (Eq. \ref{fibsyms}). Bottom: local operators in the cluster state Hamiltonian in Eq. \ref{fibexample}. Gray lines represent the graph $\textbf{G}$ on which the cluster state is defined.}
    \label{fibfig}
\end{figure}

Another class of models with subsystem symmetries are fractal SPTs (FSPTs). In a FSPT, the symmetries are Pauli $X$ operators acting on fractals on two square sublattices, $A$ and $B$. These symmetries are determined by 1D cellular automata, where each row corresponds to a time step of a cellular automaton~\cite{devakul2018fractal}. 
An example of a nontrivial~\cite{devakul2018fractal} SSPT with fractal symmetry is given by:
\begin{equation}\label{fibexample}
\begin{split}
    H=\sum_{ij}&\left[X^{(A)}_{i,j-1}Z^{(B)}_{i,j-1} Z^{(B)}_{i-1,j}Z^{(B)}_{i,j}Z^{(B)}_{i+1,j}\right. + \\ &\left.X^{(B)}_{i,j+1}Z^{(A)}_{i,j+1}
    Z^{(A)}_{i-1,j}Z^{(A)}_{i,j}Z^{(A)}_{i+1,j}\right] .
\end{split}
\end{equation}
This is also an example of a bipartite graph cluster state, with $A$ and $B$ corresponding to the two partitions of the graph. 
The symmetries and local operators of this model, which is called the Fibonacci fractal spin liquid, are depicted in Fig.~\ref{fibfig}.
We remark that they form a different symmetry group from the Sierpinski fractal introduced in Section~\ref{sec:subsyms}. 

In this work, we study \textit{first order} FSPTs in particular. First order fractal symmetries correspond to first order linear cellular automata, where each time step only depends on the time step before it. The Hamiltonian in Eq.~\eqref{fibexample} is an example of a first order FSPT, as we show below.

To describe the class of first order fractal SPTs, it is convenient to use polynomial notation. In this notation, we take a polynomial whose coefficients are in $\mathbb{F}_2$, and  act on sites corresponding to the exponents of each term with coefficient 1. For example, we define $X(x^i y^j+x^ky^l+...)$ to be $X_{i,j}X_{k,l}\times...$, and similarly for $Z$. Then, Eq.~\eqref{fibexample} becomes
\begin{equation}\label{fibstabsgeneral}
\begin{split}
    H=&\sum_{ij}\left[X^{(A)}(x^iy^j)Z^{(B)}(x^iy^j(1+yf(x)))\right.+\\
    &\left.X^{(B)}(x^iy^j)Z^{(A)}(x^iy^j(1+y^{-1}f(x^{-1})))\right],
\end{split}
\end{equation}
with $f(x) = 1+x+1/x$. 
Its symmetry generators are 
\begin{equation}\label{fibsyms}
\begin{split}
    X^{(A)}(x^i(1+f(x)y+f^2(x)y^2+...)),\\
    X^{(B)}(x^i(1+f(x^{-1})y^{-1}+f^2(x^{-1})y^{-2}+...)),
\end{split}
\end{equation}
for any $i$. Here, the coefficients of $f^k(x)$ are taken mod 2. This mod 2 multiplication is the reason why the symmetries in Eq.~\eqref{fibsyms} are fractals. Since for any polynomial $g(x)$ over $\mathbb{F}_2$, $g(x)^{2^n} = g(x^{2^n})$, the symmetries exhibit self-similarity on regions with $y$-length given by powers of 2. These symmetries are first order since the polynomial in each row is equal to the previous row multiplied by $f(x)$.

It is now easy to describe the general class of first order fractal symmetries: simply take Eq.~\ref{fibsyms}, and allow for any Laurent polynomial $f(x)$ over $\mathbb{F}_2$ \footnote{Strictly speaking, there are restrictions on $f(x)$ depending on the boundary conditions used. This subtlety is briefly discussed in the appendix.}. Every first order FSPT has symmetry group in Eq. \ref{fibsyms} corresponding to some polynomial $f(x)$. For example, the Sierpinski fractal from Section \ref{sec:subsyms} is given by the polynomial $f(x)=1+x$. There are many fractal SPT phases for the same symmetry group. A classification of all of these phases, based on the twist phase, was given in Ref.~\onlinecite{Devakul2018}.

In this work, we study a subclass of first order FSPTs as examples: cluster states with two qubits per unit cell. 
Since there are two sublattices, the minimal unit cell under which the symmetries can act nontrivially contains two qubits. For the rest of the work, we use the somewhat more compact terminology \textit{FSPT cluster state} to refer to cluster states that are first order FSPTs with two qubits per unit cell. 
Clearly, Eq.~\eqref{fibexample} is an example of an FSPT cluster state\footnote{The two qubit unit cell fractal SPT Hamiltonians are classified in Ref.~\onlinecite{devakul2018fractal}, along with more general models. For further details, see Appendix~\ref{appendixfractal}}.

To find ERG fixed points for the FSPT cluster state models, we first find symmetric LUCs for general first order fractal symmetries. It is again convenient to utilize polynomial notation. Recall that the sets $Q_A$ and $Q_B$, introduced in Section \ref{sec:cluster} to define symmetry-preserving LUCs, are local. Hence, they can be written as finite degree polynomials $Q_A(x,y)$ and $Q_B(x,y)$. We find that these polynomials can be further decomposed into ${Q_A(x,y) = c(x,y)P_A(x,y)}$ and ${Q_B(x,y) = d(x,y)P_B(x,y)}$, where $c$ and $d$ are arbitrary finite degree polynomials, and 
\begin{equation}\label{fractallucs}
    P_A(x,y)= 1 + y^{-1}f(x^{-1}),\quad P_B(x,y)=1+yf(x) \, . 
\end{equation}
With this general functional form of $Q_A$ and $Q_B$, we then search for specific $h$ and $g$ to construct symmetric local circuits for the ERG flow. The exact forms of $h$ and $g$ used for the FSPT cluster state models are given in Section 4 of the Mathematica notebook. 

Using Eq. \ref{fractallucs}, we find that there are only three unique fixed points for all FSPT cluster states, for general polynomials. The first fixed point, $H_A$ is self-bifurcating, while the other two, $H_+$ and $H_-$ are quotient-bifurcating fixed points. 
\begin{align}
H_{\pm}(a) &\simeq H_{\pm}(2a) + H_A(2a) ,
\\
H_{A}(a) &\simeq H_A(2a) + H_A(2a).
\end{align}

\begin{figure}[t!]
    \centering
    \includegraphics[width=\columnwidth]{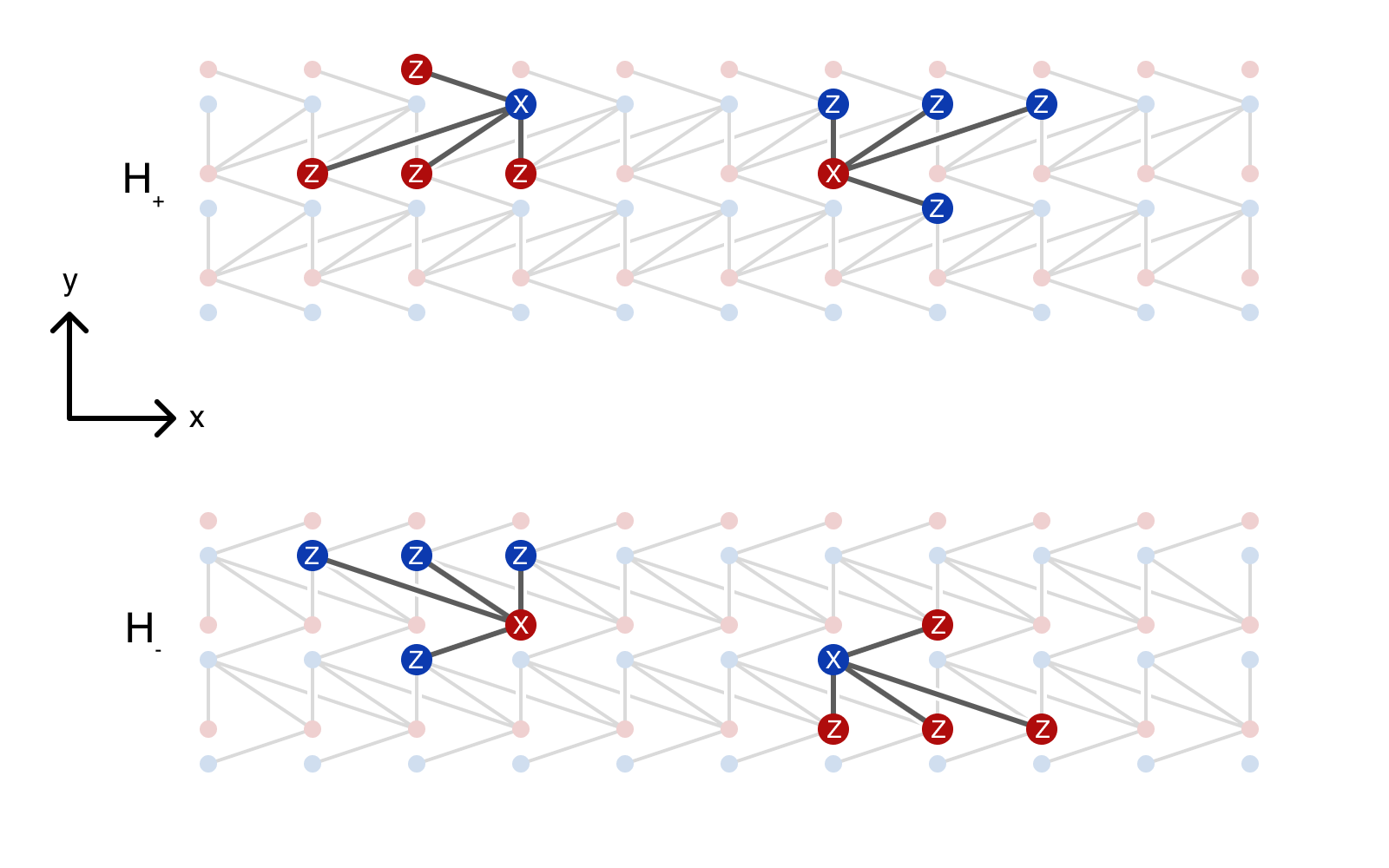}
    \caption{Local operators of the fixed points $H_+$ and $H_-$ corresponding to the polynomial $1+x+1/x$.}
    \label{fibfixedpointsfig}
\end{figure}

To define $H_A$, we simply generalize Eq.~\eqref{fibstabsgeneral} to arbitrary $f(x)$. 
This model has been discussed before in the literature; it is simply the first order fractal spin liquid (FSL) from Ref.~\onlinecite{yoshida2013exotic} with $g(x)=0$, up to a Hadamard operation on half the qubits. 
The Hamiltonians $H_+$ and $H_-$ are similar, given by
\begin{equation}\label{hplusminus}
\begin{split}
    H_{\pm}=&\sum_{ij}\left[X^{(A)}(x^iy^j)Z^{(B)}(x^{i\pm 1}y^j(1+yf(x)))\right.+\\
    &\left.X^{(B)}(x^{i}y^j)Z^{(A)}(x^{i \mp 1}y^j(1+f(1/x)/y))\right]
\end{split} 
\end{equation}
which correspond to shifted first order FSLs. 
They are related to $H_A$ by shifting one of the sublattices by one site, either to the left or to the right. We remark that this operation is not an allowed SSPT phase equivalence. 
Fig.~\ref{fibfixedpointsfig} gives an example of $H_{\pm}$ for the polynomial $1+x+1/x$ discussed earlier. The details of the ERG flow to reach these fixed points is provided in Appendix \ref{appendixfractal}.  
Each of the Hamiltonians $H_A,H_+,H_-$ have distinct twist phases, and hence are in inequivalent SSPT phases~\footnote{Redefinitions of the lattice are not allowed in SSPT phase equivalence, since they are not symmetric local unitaries. However, if such redefinitions are added to the phase equivalences, the three phases become equivalent.}. 
In the appendix we further demonstrate that $H_{\pm}$ cannot self-bifurcate. 

\begin{figure}[t!]
    \centering
    \includegraphics[width=\columnwidth]{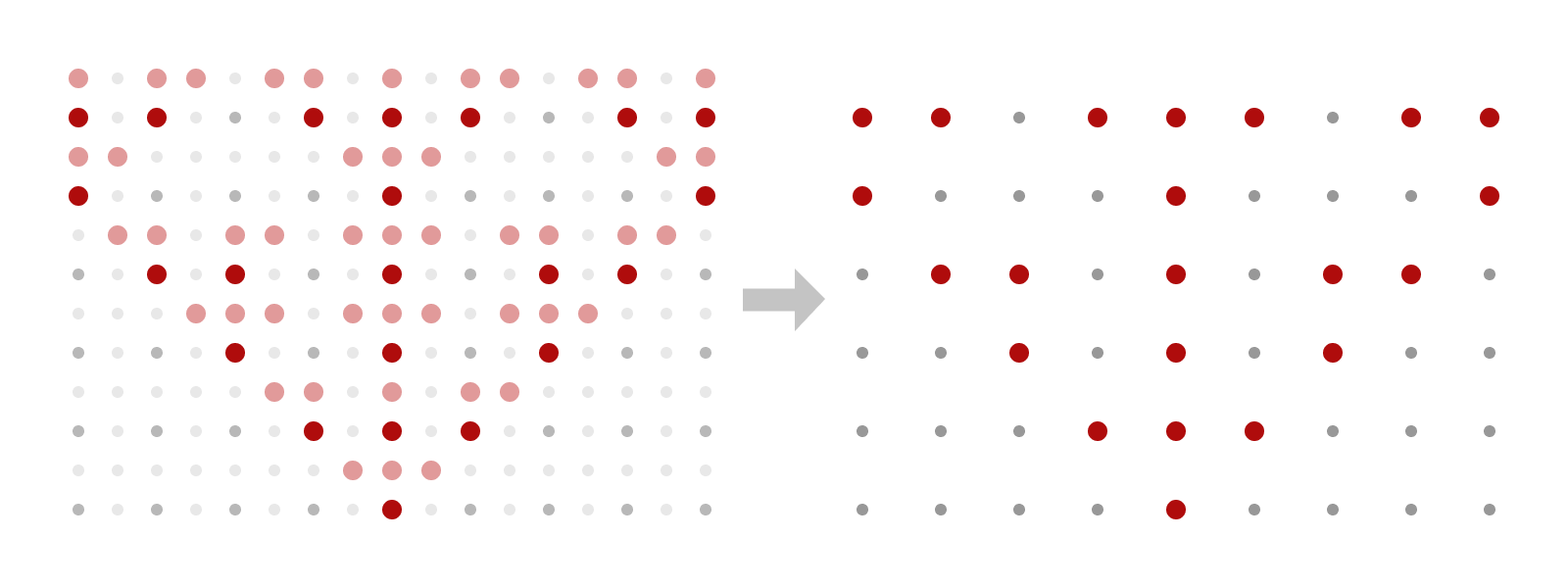}
    \caption{Coarse-graining a fractal symmetry operator for $f(x)=1+x+1/x$. While individual symmetry operators are not always preserved as in this case, the entire symmetry group is.}
    \label{fractalsymbifurcationfig}
\end{figure}

In Section \ref{sec:ssergdef}, we required that the symmetry group self-bifurcates as well as the Hamiltonian. It may not be obvious from Eq. \ref{fibsyms} that this is the case; however, due to the self-similarity of polynomials in $\mathbb{F}_2$, this is true, as long as we coarse-grain by a factor of two~\cite{yoshida2013exotic}. If we choose any one site per coarse-grained unit cell, then $G_i(2a)$ acting on that site is always equivalent to $G(a)$. An example of this is shown in Fig.~\ref{fractalsymbifurcationfig}. While individual generators are not always preserved, the symmetry group restricted onto each sublattice is always equivalent to the original group, as defined in Eq.~\ref{subsystemequiv}. 

The three fixed points we have described above allow for a consolidation of the existing classification. We remark that even if we restrict to a two qubit unit cell, there exist an infinite number of phases corresponding to a given polynomial~\cite{Devakul2018}. These phases require increasingly nonlocal Hamiltonians, in that they have local operators with larger and larger support. However, using the ERG, these simply reduce to copies of the three phases given in Eq.~\eqref{fibstabsgeneral} and Eq.~\eqref{hplusminus}. The locality of these phases (the spread of local operators on the coarse-grained lattice) depends only on the polynomial $f(x)$.\par
These results give us an alternate way to study fractal phases: reducing them to a small set of highly local models. 
While this does not yet give a complete ERG based classification for all FSPT phases,
it should be possible to extend our approach to more general fractal Hamiltonians in the future.

\section{Discussion \& conclusion}
\label{sec:Discussion}

In this work, we introduced the subsystem symmetry-preserving entanglement renormalization group and found nontrivial symmetric gapped bifurcating fixed points in two dimensions. 
We outlined an approach for finding these fixed points within subsystem SPT phases. 
Our approach utilized twist phases to constrain potential solutions, combined with properties of cluster state Hamiltonians on bipartite graphs to explicitly generate symmetric local unitary circuits. 
We applied this method to examples of fractal and linear subsystem symmetry protected phases in two dimensions. In both cases, we found self- and quotient-bifurcating fixed points, similar to those of fracton topological orders. 

For the linear cluster state, we found quotient bifurcation, where the Hamiltonian splits into a copy of itself and a separate, self-bifurcating Hamiltonian. 
For the general class of FSPT cluster states, we found three possible fixed points, all of which are related to the FSPTs defined by Yoshida's fractal spin liquids~\cite{yoshida2013exotic} up to shifts of one of the two sublattices. These include both self- and quotient-bifurcating fixed points.
To the best of our knowledge, no such examples of symmetric gapped bifurcating fixed points in two dimensions have previously appeared in the literature.

Our work has potential implications for the classification of SSPT phases: 
first of all, we expect that SSPERG flows can be applied to reduce the classification from Ref.~\onlinecite{Devakul2018} by removing the length scale at the cost of increasing the size of the on-site symmetry group via coarse-graining. 
This could lead to a classification that is finite for any fixed finite symmetry group, without having to introduce a length scale cutoff, similar to the known classifications of global SPTs~\cite{chen2013symmetry}.

Secondly, the discovery of self- and quotient-bifurcating SSPTs in this work inspires the definition of a looser quotient equivalence relation on SSPT phases by modding out self-bifurcating SSPTs, following a similar definition that has been introduced for fracton phases~\cite{shirley2017fracton,Dua2019b}. 
The purpose of establishing this definition 
is that quotient-bifurcating SSPERG fixed points provide representatives for quotient equivalence classes of SSPT phases. 
This is a generalization of the more familiar classification of SPT phases via representative gapped fixed points under symmetry-preserving ERG~\cite{gu2009tensor,Chen2011}. 
In particular, under this definition, our linear SSPT example and the fractal SSPT examples $H_\pm$ are quotient-bifurcating fixed point representatives of nontrivial equivalence classes, while the fractal spin liquid $H_A$, which self-bifurcates, is in the ``trivial'' equivalence class.


To date there have been several attempts to define a looser equivalence relation on linear SSPT phases that quotients out phases given by decoupled 1D SPT layers~\cite{subsystemphaserel,Devakul2018,Shirley2019d,Devakul2020}. 
This is indeed achieved by our proposed definition of quotient equivalence classes, as a stack of 1D SPTs is trivially a self-bifurcating fixed-point. 
We remark, however, that certain linear SSPTs with intersecting symmetries that act via the same on-site operators cannot lead to B models that are trivial decoupled 1D SPT layers under bifurcating SSPERG. 
In particular, the twist phases of the square lattice cluster model are inconsistent with a B model consisting of decoupled 1D SPTs under bifurcating SSPERG~\cite{subsystemphaserel}.

We note that symmetric locality-preserving unitaries that are not local unitaries, such as partial translations, are not included as a phase equivalence relation. In particular these can change, and even trivialize, nontrivial SSPT phases. 
This is familiar from the study of 1D SPT phases, where a partial translation can map a nontrivial SPT state to a trivial one. 
For example consider a system consisting of pairs of spin-$\frac{1}{2}$ particles on each site $s$, labelled $a_s,b_s$. 
Take the nontrivial SPT state, under the tensor product action of $SU(2)$ on each pair of spins, given by maximally entangled singlet states between $b_s,a_{s+1},$ on neighboring sites. Applying a partial translation to the $b_s$ spins results in a trivial SPT state, given by a tensor product of singlet states on each site between spins $a_s,b_s$. 
It remains unclear whether a subclass of these more general locality-preserving transformations can be included to allow for looser phase equivalence without trivializing all SSPT phases. 
In particular, a partial lattice translation relates the $H_A$ fractal SSPT to the Hamiltonians $H_\pm$, which lie in distinct SSPT phases.

During this work we ran across a number of interesting unresolved questions:
\begin{itemize}
    \item 
Are the twist phase invariants complete, or can they be further refined? 
\item
Can two states in the same SSPT phase, which is technically defined by equivalence classes of gapped symmetric Hamiltonians connected by adiabatic gapped symmetry-preserving paths, be (approximately) related by a symmetry-preserving local unitary transformation, up to the addition of trivial symmetric degrees of freedom? (The converse is obvious).
\item 
To what extent are bifurcating SSPERG fixed points unique? What is the appropriate equivalence relation between them?
\item 
How do our results on the bifurcating ERG behaviours of SSPTs relate to their use as universal resources for MBQC? 
\end{itemize}
We leave these questions to future work. 

\acknowledgments
AD thanks Meng Cheng and Wilbur Shirley for useful discussions and acknowledges support from the Simons Collaboration on Ultra-Quantum Matter. DJW acknowledges support from the Simons foundation. JSM acknowledges support from the National Science Foundation Graduate Research Fellowship under Grant Number DGE-1656518.

\bibliographystyle{apsrev4-1}
\bibliography{DomBib}

\appendix

\section{Polynomial notation}
\label{app:polynomial}

In this appendix, we review the polynomial notation for Pauli operators~\cite{haah2013commuting,Haah2013}, a generalization of the notation we used previously in Eq. \ref{fibstabsgeneral}. We also describe coarse-graining and the calculation of twist phases in this notation. We have provided the code in a supplemental Mathematica notebook to perform these operations.

\begin{figure}
    \centering
    \includegraphics[width=\columnwidth]{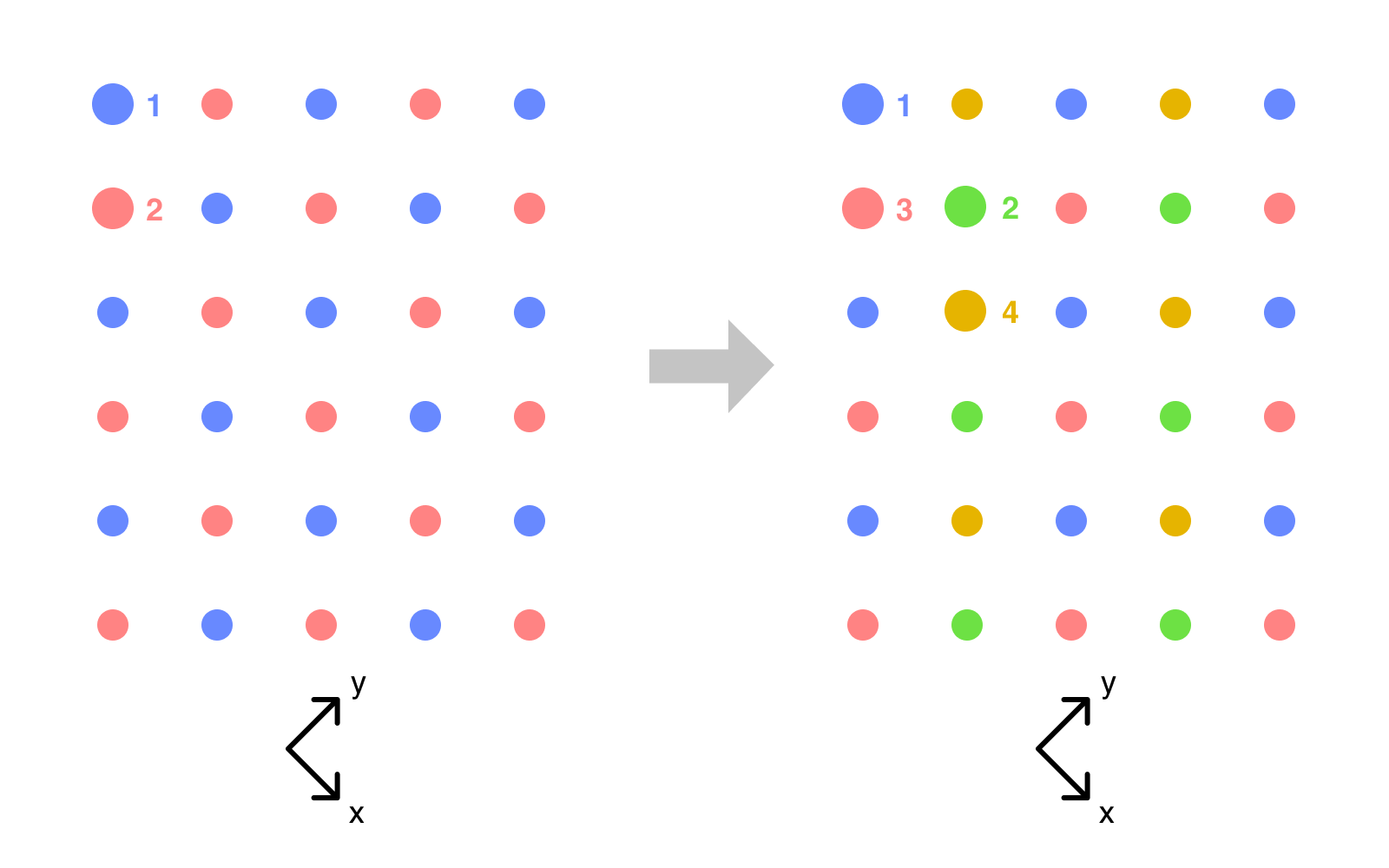}
    \caption{Labelling of qubits in the unit cell for the linear cluster model. Left: coordinates in Eq. \ref{linearclusterpoly}. Right: coordinates after coarse-graining in $x$ by a factor of 2 (Eq. \ref{linearclustercgx}).}
    \label{coordlabelsfig}
\end{figure}

In our previous notation, we could describe any Pauli operator (up to  a phase) via 
\begin{align}
X^{(a)}(f_1(x,y))Z^{(a)}(f_2(x,y))X^{(b)}(f_3(x,y))Z^{(b)}(f_4(x,y))...
\end{align}
Here, $(a),(b),...$ represent sites within a unit cell, and $f_i(x,y)$ are Laurent polynomials in $\mathbb{F}_2$. 
For example, in Eq. \ref{fibstabsgeneral}, there are two sites, $A$ and $B$, each with one qubit per cell. The sites in the unit cell may correspond to the partition of the cluster state graph $\textbf{G}$, as in that example, or they may be enumerated according to some other scheme. Following previous work due to Haah~\cite{haah2013commuting,Haah2013}, we now compress this notation into a single vector,
\begin{align}\label{polynotationex}
    \begin{pmatrix}
    f_1(x,y)\\
    f_3(x,y)\\
    \vdots\\
    f_2(x,y)\\
    f_4(x,y)\\
    \vdots
    \end{pmatrix}.
\end{align}
The top entries of Eq. \ref{polynotationex} are the $X$ operators on each lattice site, and the bottom entries are the $Z$ operators. We can use this notation to write out an entire local Hamiltonian. We write the Hamiltonian as a matrix, where each column is one of the translationally invariant local terms, represented as in Eq. \ref{polynotationex}. As an example, consider the 2D linear cluster state in Eq.\ref{2dcluster1}, with lattice sites labelled as in Fig.  \ref{coordlabelsfig}. Then, the Hamiltonian may be written
\begin{align}\label{linearclusterpoly}
    H_{lin}=\begin{pmatrix}
    0 & 1\\
    1 & 0\\
    1+x+\bar{y}+x\bar{y} & 0\\
    0 & 1+\bar{x}+y+\bar{x}y
    \end{pmatrix}.
\end{align}
For additional compactness, we have written $x^{-1}$ and $y^{-1}$ as $\bar{x}$ and $\bar{y}$, respectively. We can write the symmetry group in a similar way,
\begin{align}\label{linearclusterpolysym}
    \begin{pmatrix}
        S(x) & S(y) & 0 & 0 \\
        0 & 0 & S(x) & S(y) \\
        0 & 0 & 0 & 0 \\
        0 & 0 & 0 & 0 \\
    \end{pmatrix},\quad S(t) = 1+t+t^2+...
\end{align}
Here, this notation means that the symmetry group is entirely generated by translations of Eq. \ref{linearclusterpolysym}. The operation of coarse-graining can also be written in this notation~\cite{Haah2013,haah2014bifurcation}. When we double the lattice spacing in one direction (say $x$), we do two things. First, we enlarge the unit cell, which doubles the indices of the vector in Eq. \ref{polynotationex}. Second, each translationally invariant operator becomes two translationally invariant operators on the coarse-grained lattice, corresponding to translations by even and odd amounts along $x$. We again return to the example in Eq. \ref{linearclusterpoly}. Doubling the unit cell along the $x$ direction, this becomes
\begin{align}\label{linearclustercgx}
H=\begin{pmatrix}
0 & 0 & 1 & 0 \\
0 & 0 & 0 & 1 \\
1 & 0 & 0 & 0 \\
0 & 1 & 0 & 0 \\
1+\bar{y} & x + x\bar{y} & 0 & 0 \\
1 + \bar{y} & 1+ \bar{y} & 0 & 0 \\
0 & 0 & 1+y & 1+y \\
0 & 0 & \bar{x}+\bar{x}y & 1+y
\end{pmatrix}.
\end{align}
The coarse-grained unit cell for this Hamiltonian is shown in Fig. \ref{coordlabelsfig}. In general, when we coarse-grain along $x$, we simply replace the variables in a vector with matrices:
\begin{align}\label{xcg}
    x \rightarrow \begin{pmatrix}
    0 & x \\
    1 & 0
    \end{pmatrix},\quad
    y \rightarrow \begin{pmatrix}
    y & 0 \\
    0 & y
    \end{pmatrix}.
\end{align}
We perform a similar transformation to coarse-grain along $y$. Note that both of these coarse-graining transformations are just one possible labelling of the coarse-grained qubits. We may, for example, permute rows and columns in Eq. \ref{xcg} for an equivalent, but different, labelling of the coarse-grained qubits. \par
Finally, we may use the polynomial notation to compactly write the twist phases. An $L\times L$ SSPT has order $L$ symmetries, and thus $L^2$ twist phases. However, most of these symmetries are just translations of each other. By writing out the symmetry group as in Eq. \ref{linearclusterpolysym}, we define a local symmetry group on each unit cell. We label this local group $S^{(1)},S^{(2)},...,S^{(n)}$. Then, we may write any member of the entire symmetry group as $S^{(k)}(x^iy^j)$. This notation means we take the $k$th element of the local group and translate by $i$ in the $x$ direction and $j$ in the $y$ direction.\par
We can now define the \textit{twist phase matrix} $W^{(x)}$. $W^{(x)}$ is a $n\times n$ matrix where $W^{(x)}_{ij} = \sum_{k}c^{(ij)}_ky^k$, and 
\begin{align}
    c^{(ij)}_k=\begin{cases}
    0, & \Omega^{(x)}(S^{(i)}(y^k),S^{(j)}(1))=1 \\
    1, & \Omega^{(x)}(S^{(i)}(y^k),S^{(j)}(1))=-1
    \end{cases},
\end{align}
where $\Omega^{(x)}$ is the twist phase calculated along an $x$ cut, defined in Eq. \ref{twistphasedef}. We also define a twist phase matrix $W^{(y)}$ for cuts made along the $y$ direction, with $W^{(y)}_{ij} = \sum_{k}b^{(ij)}_kx^k$, and
\begin{align}
    b^{(ij)}_k=\begin{cases}
    0, & \Omega^{(y)}(S^{(i)}(x^k),S^{(j)}(1))=1 \\
    1, & \Omega^{(y)}(S^{(i)}(x^k),S^{(j)}(1))=-1
    \end{cases}.
\end{align}
Notice that, in either case, the polynomials $b^{(ij)}$ and $c^{(ij)}$ are only polynomials in one variable. This is because there are not extensively many subsystem symmetries; in other words, each operator in our local basis only gives an independent set of operators if translated along one direction. We may then choose a basis such that each twist phase matrix is only a function of one variable. \par
As a final example, we calculate the twist phases for the symmetries defined in Eq. \ref{linearclusterpolysym} and the Hamiltonian in Eq. \ref{linearclusterpoly}. Two of these symmetries are lines in the $x$ direction, and two are lines in the $y$ direction.\par
If we make cuts along constant $y$, then the $x$-type symmetries are not truncated at all, and therefore have no twist phases with other symmetries. Meanwhile, the $y$-type symmetries only have nontrivial twist phases with their nearest neighbors,
\begin{align}\label{linearclusterytwist}
    W^{(y)}=\begin{pmatrix}
    0 & 0 & 0 & 0 \\
    0 & 0 & 0 & 1+x \\
    0 & 0 & 0 & 0 \\
    0 & 1+\bar{x} & 0 & 0
    \end{pmatrix}.
\end{align}
Similarly, we can show
\begin{align}\label{linearclusterxtwist}
    W^{(x)}=\begin{pmatrix}
    0 & 0 & 1+\bar{y} & 0 \\
    0 & 0 & 0 & 0 \\
    1+y & 0 & 0 & 0 \\
    0 & 0 & 0 & 0 
    \end{pmatrix}.
\end{align}

\section{ERG for the linear cluster state}
\label{linearappendix}

\begin{figure}[t!]
    \centering
    \includegraphics[width=\columnwidth]{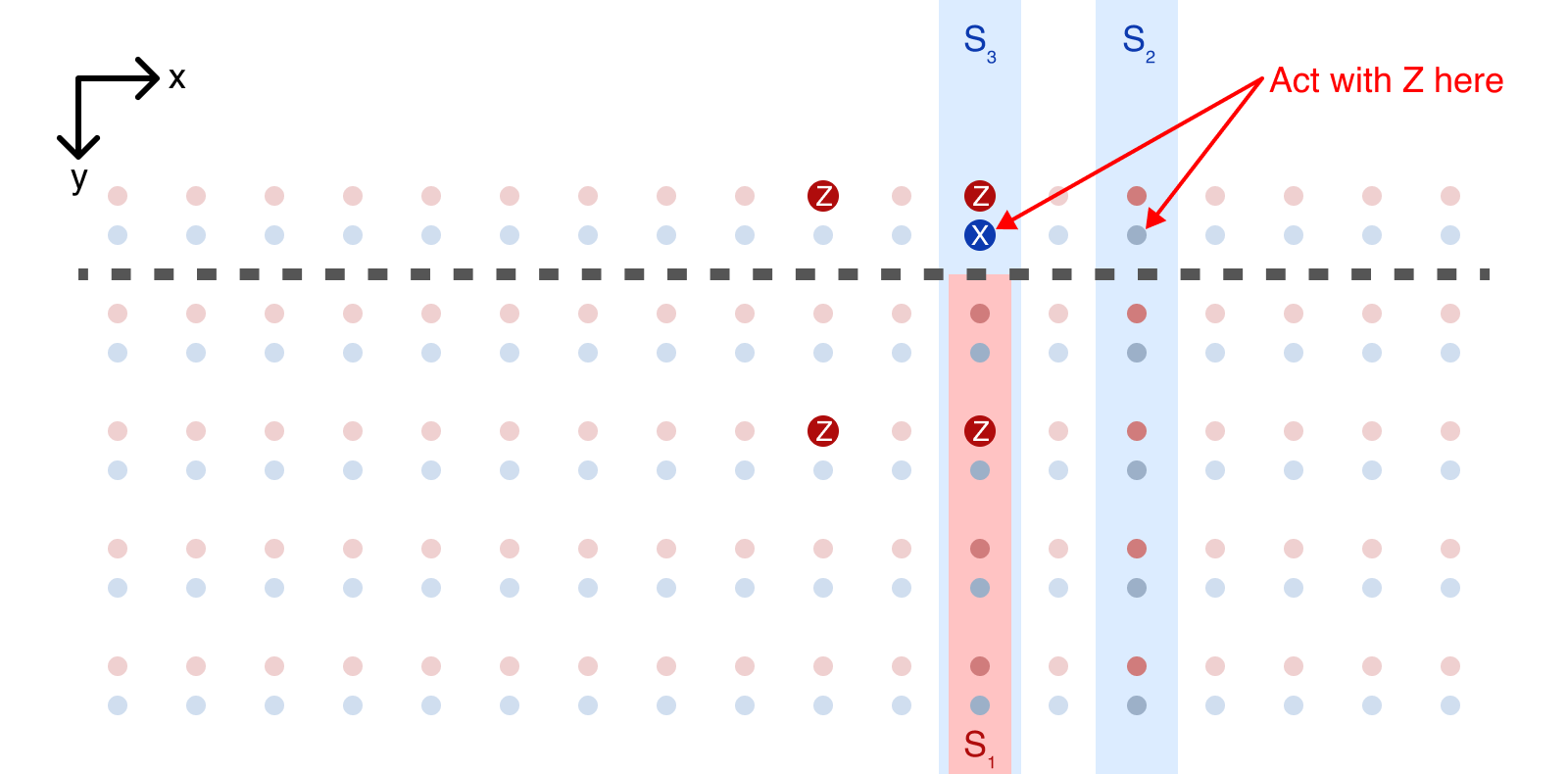}
    \caption{Calculating twist phase for one copy of the linear cluster state on a coarse-grained lattice. We start with a truncated symmetry, $S_1$. It anticommutes with two local terms in $H_{lin}(2a)$ (one of these is pictured). To get the same action, we apply a Pauli $Z$ operator to two sites, which anticommutes with the symmetries $S_2$ and $S_3$. These symmetries are always two sites apart on the original lattice.}
    \label{lineartwistphasefig}
\end{figure}

Using the tools and definitions of the previous section, we now study the ERG of the 2D linear cluster state. As previously mentioned, the linear cluster Hamiltonian (Eq. \ref{linearclusterpoly}) has twist phase matrices given by Eq. \ref{linearclusterytwist} and Eq. \ref{linearclusterxtwist}. After coarse-graining, first along $y$ and then along $x$, the symmetries may be written
\begin{align}\label{cglinearsyms}
\begin{pmatrix}
S(x) & 0 & S(y) & 0 & 0 & 0 & 0 & 0 \\
0 & S(x) & S(y) & 0 & 0 & 0 & 0 & 0 \\
S(x) & 0 & 0 & S(y) & 0 & 0 & 0 & 0 \\
0 & S(x) & 0 & S(y) & 0 & 0 & 0 & 0 \\
0 & 0 & 0 & 0 & S(x) & 0 & S(y) & 0 \\
0 & 0 & 0 & 0 & 0 & S(x) & S(y) & 0 \\
0 & 0 & 0 & 0 & S(x) & 0 & 0 & S(y) \\
0 & 0 & 0 & 0 & 0 & S(x) & 0 & S(y) \\
0 & 0 & 0 & 0 & 0 & 0 & 0 & 0 \\
\multicolumn{8}{c}{$\vdots$} \\
0 & 0 & 0 & 0 & 0 & 0 & 0 & 0 
\end{pmatrix}.
\end{align}
Notice that we have dropped half of the columns that appear when carrying out the transformations in Eq. \ref{xcg}. This is because $S(y)$ and $S(x)$ are unchanged by translations by $y$ and $x$, respectively. Therefore, some columns (symmetry operators) are redundant due to equivalence under translation. 
This reduction during coarse-graining is a general property of subsystem symmetries. 

In the basis given in Eq. \ref{cglinearsyms}, the twist phases are 
\begin{align}\label{lineartwistphasecg}
    W^{(y)}=\begin{pmatrix}
0 & 0 & 0 & 0 & 0 & 0 & 0 & 0 \\
0 & 0 & 0 & 0 & 0 & 0 & 0 & 0 \\
0 & 0 & 0 & 0 & 0 & 0 & 1 & 1 \\
0 & 0 & 0 & 0 & 0 & 0 & x & 1 \\
0 & 0 & 0 & 0 & 0 & 0 & 0 & 0 \\
0 & 0 & 0 & 0 & 0 & 0 & 0 & 0 \\
0 & 0 & 1 & \bar{x} & 0 & 0 & 0 & 0 \\
0 & 0 & 1 & 1 & 0 & 0 & 0 & 0
\end{pmatrix},
\end{align}
and
\begin{align}
    W^{(x)}=
\begin{pmatrix}
0 & 0 & 0 & 0 & 1 & \bar{y} & 0 & 0 \\
0 & 0 & 0 & 0 & 1 & 1 & 0 & 0 \\
0 & 0 & 0 & 0 & 0 & 0 & 0 & 0 \\
0 & 0 & 0 & 0 & 0 & 0 & 0 & 0 \\
1 & 1 & 0 & 0 & 0 & 0 & 0 & 0 \\
y & 1 & 0 & 0 & 0 & 0 & 0 & 0 \\
0 & 0 & 0 & 0 & 0 & 0 & 0 & 0 \\
0 & 0 & 0 & 0 & 0 & 0 & 0 & 0 
\end{pmatrix}.
\end{align}
Now, consider the following model on the coarse-grained unit cell:
\begin{widetext}
\begin{align}\label{linearclusteraftergates}
    H_2=\begin{pmatrix}
    0 & 0 & 0 & 0 & 1 & 0 & 0 & 0\\
0 & 0 & 0 & 0 & 0 & 0 & 0 & 1\\
0 & 0 & 0 & 0 & 0 & 1 & 0 & 0\\
0 & 0 & 0 & 0 & 0 & 0 & 1 & 0 \\
0 & 1 & 0 & 0 & 0 & 0 & 0 & 0 \\
0 & 0 & 1 & 0 & 0 & 0 & 0 & 0\\
1 & 0 & 0 & 0 & 0 & 0 & 0 & 0\\
0 & 0 & 0 & 1 & 0 & 0 & 0 & 0\\
0 & 0 & 1+y & 1+y & 0 & 0 & 0 & 0 \\
1+x+\bar{y}+x\bar{y} & 0 & 0 & 0 & 0 & 0 & 0 & 0 \\
0 & 1+\bar{x} & \bar{x}+y & 1+y & 0 & 0 & 0 & 0\\
0 & 1+\bar{x} & 1+\bar{x} & 0 & 0 & 0 & 0 & 0\\
0 & 0 & 0 & 0 & 0 & 1+x & 1+x & 0\\
0 & 0 & 0 & 0 & 1+\bar{y} & x+\bar{y} & 1+x & 0\\
0 & 0 & 0 & 0 & 0 & 0 & 0 & 1+\bar{x}+y+\bar{x}y \\
0 & 0 & 0 & 0 & 1+\bar{y} & 1+\bar{y} & 0 & 0\\
\end{pmatrix}.
\end{align}
\end{widetext}
$H_2$ can be split into two Hamiltonians with disjoint support. The first Hamiltonian is supported on the second and the seventh qubits, and is a copy of the original Hamiltonian, (Eq. \ref{linearclusterpoly}). This is $H_{lin}(2a)$. The second, $H_B(2a)$, has support on the other 6 qubits. The combination of both of these models, Eq. \ref{linearclusteraftergates}, satisfies the twist phases in Eq. \ref{lineartwistphasecg}, as shown in Section 3 of the Mathematica notebook. In the notebook, we further show that $H_B(2a)$ is self-bifurcating, and write the local circuits required to achieve bifurcation for both models.

We now give an argument that self-bifurcation of $H_{lin}$ is impossible. 
We assume the contrary, and suppose that we have two copies of $H_{lin}(2a)$, and that these, combined, satisfy the twist phases in Eq. \ref{lineartwistphasecg}. 
Then, we consider one of these copies on the original lattice, and make a cut at constant $y$ (Fig. \ref{lineartwistphasefig}). A given truncated vertical symmetry anticommutes with two local terms in the copy, each two lattice spacings apart in the $x$ direction. An edge operator with the same action is two $Z$ operators, also two lattice spacings apart in $x$. The other copy has similar edge operators, and the product of these operators defines the twist phases. However, since both copies give $Z$ operators that are two lattice spacings apart, we cannot get twist phases between only a symmetry and its immediate neighbors, simply as a consequence of parity. Therefore, the linear cluster state cannot self-bifurcate.

\section{ERG for fractal SPT cluster states}
\label{appendixfractal}

In this appendix, we write down the most general form for FSPT cluster states, as defined in Section \ref{sec:fractal}, and then describe their flow under  ERG. Finally, we show that two of our fixed points, $H_\pm$ (Eq. \ref{hplusminus}), cannot self-bifurcate. \par 
In order to write down the general FSPT form, it is important to mention some facts about boundary conditions that we have previously avoided in the main text.

The symmetries in Eq. \ref{fibsyms} are, strictly speaking, only symmetries of Eq. \ref{fibstabsgeneral} under certain periodic boundary conditions and with certain polynomials. In particular, under periodic boundary conditions, we set $x^{2^{l_x}}=y^{2^{l_y}}=1$, for some $l_x$ and $l_y$. Then, Eq. \ref{fibsyms} requires that $f(x^{2^{l_y}})=1$ to be well-defined. This is not a particularly restrictive condition, as we can satisfy it by setting $l_y > l_x$ and using any $f(x)$ such that $f(1)=1$. Any polynomial with an odd number of terms, such as our example $1+x+1/x$ works.\par
For arbitrary polynomials, we can use open boundary conditions to get well-defined models, but we do not discuss this possibility in this work. Instead, we define fractal symmetries to be
 \begin{align}\label{fractalsymspoly}
     \begin{pmatrix}
     \mathcal{F}(x,y) & 0 \\
     0 & \mathcal{F}(\bar{x},\bar{y}) \\
     0 & 0 \\
     0 & 0
     \end{pmatrix},
 \end{align}
 where
 \begin{align}
     \mathcal{F}(\bar{x},\bar{y})=\sum_{i=0}^{2^{l^y}}y^if(x)^i.
 \end{align}
We now write down an arbitrary FSPT cluster state.  For a given polynomial $f(x)$, such an SPT can be written
\begin{align}\label{higherfractal}
    H=\begin{pmatrix}
    0 & 1 \\
    1 & 0 \\
    y^l w(x)(1+yf(x)) & 0 \\
    0 & \bar{y}^l w(\bar{x})(1+\bar{y}f(\bar{x}))
    \end{pmatrix},
\end{align}
 where $w(x)$ is an arbitrary polynomial and $l$ is an arbitrary integer. Our three fixed points all correspond to Eq. \ref{higherfractal} with $l=0$ and different values of $w(x)$. In $H_A$, $w(x)=1$ and in $H_\pm$, $w(x)=x^{\pm 1}$.\par
 
 As an important aside, these fixed point Hamiltonians can also be written in the $\textbf{K}$-matrix notation used in the twist-phase classification of Ref. \onlinecite{Devakul2018}. In this notation, the twist phases are calculated using the local basis
 \begin{align}
     \begin{pmatrix}
     \mathcal{F}(x,y) & 0 \\
     0 & y^l\mathcal{F}(\bar{x},\bar{y}) \\
     0 & 0 \\
     0 & 0
     \end{pmatrix}.
 \end{align}
 Note the factor of $y^l$ here. The twist phase classification uses a basis of matrices $\textbf{K}^{(k,m)}$ to describe every possible set of twist phases. The matrices $\textbf{K}^{(0,m)}$ correspond to all of the translationally invariant models, which we study in this work. More precisely, in our notation, $\textbf{K}^{(0,m)}$ corresponds to $W^{(y)}_{12}=x^m$ and $W^{(y)}_{11}=W^{(y)}_{22}=0$. As we show below by calculating twist phases, $H_A$ corresponds to $\textbf{K}^{(0,0)}$, and $H_\pm$ corresponds to $\textbf{K}^{(0,\pm 1)}$.
 
We now calculate the $y$ twist phases of Eq. \ref{higherfractal}. First, we solve the $l=0$ case. The $l\neq 0$ case does not add significant complications; if $l\neq 0$, the factor $y^l$ effectively shifts one of the sublattices by $l$ cells in the $y$ direction. We can remove it by shifting it back with a unitary operator, call it $P^{-l}$. $P^{-l}$ is not a local unitary circuit; however, $P^{l}UP^{-l}$ is, as long as $U$ is a local unitary circuit. We can therefore solve the $l=0$ case, and then shift our result.\par
 In the $l=0$ case, the $y$ twist phases of Eq. \ref{higherfractal} are given by
\begin{align}\label{fractaltwistphasesbeforecg}
     W^{(y)}=\begin{pmatrix}
        0 & w(x) \\
        w(\bar{x}) & 0
     \end{pmatrix}.
\end{align}
In the fractal SPT, only the $y$ twist phases are needed. The reason is due to subsystem symmetry: translations of the fractals along the $x$ and $y$ direction are not independent. It can be shown that this results in the $x$ and $y$ twist phases also not being independent~\cite{Devakul2018}. We choose to calculate the $y$ twist phases because cutting the symmetries along $y$ does not depend on our choice of polynomial $f(x)$.\par
We now coarse-grain, first along $y$, and then along $x$. General polynomials, such as $w(x)$ and $f(x)$, have both even and odd terms, which behave differently under coarse-graining. For any polynomial, we may write
\begin{align}\label{polycoarsegrain}
    w(x) \rightarrow \begin{pmatrix}
    u(x) & xv(x) \\
    v(x) & u(x)
    \end{pmatrix}.
\end{align}
Using the fact that $\mathcal{F}(x,y)=(1+yf(x))\mathcal{F}(x^2,y^2)$, we can coarse-grain Eq. \ref{fractalsymspoly} along $y$:
\begin{widetext}
\begin{align} \label{fractalcgy}
\begin{pmatrix}
    \mathcal{F}(x^2,y) & yf(x)\mathcal{F}(x^2, y) & 0 & 0 \\
    f(x)\mathcal{F}(x^2,y) & \mathcal{F}(x^2,y) & 0 & 0 \\
    0 & 0 & \mathcal{F}(\bar{x}^2,\bar{y}) & f(\bar{x})\mathcal{F}(\bar{x}^2,\bar{y}) \\
    0 & 0 & \bar{y}f(\bar{x})\mathcal{F}(\bar{x}^2,\bar{y}) & \mathcal{F}(\bar{x}^2,\bar{y}) \\
    0 & 0 & 0 & 0 \\
    0 & 0 & 0 & 0 \\
    0 & 0 & 0 & 0 \\
    0 & 0 & 0 & 0
\end{pmatrix}.
\end{align}
\end{widetext}
Now, periodic boundary conditions imply that $yf(x^2)\mathcal{F}(x^2,y)=\mathcal{F}(x^2,y)$. This in turn implies that only two of the columns in are independent. This, of course, is due to fractal symmetries being subsystem symmetries. Because of this, we drop the second and fourth symmetries from Eq. \ref{fractalcgy}. In this basis, the twist phases are again given by Eq. \ref{fractaltwistphasesbeforecg}. We may then coarse-grain along $x$, which gives us
\begin{align}\label{fractaltwistphasecg}
    W^{(y)}=\begin{pmatrix}
    0 & 0 & u(x) & v(x) \\
    0 & 0 & xv(x) & u(x) \\
    u(\bar{x}) & \bar{x}v(\bar{x}) & 0 & 0 \\
    v(\bar{x}) & u(\bar{x}) & 0 & 0
    \end{pmatrix}.
\end{align}
In Section 2 of the Mathematica notebook, we show that these twist phases can be satisfied by four copies of the original model (Eq. \ref{higherfractal}), with different values of $w(x)$. Two copies have $w(x)=u(x)$, a third has $w(x)=xv(x)$, and the last has $w(x)=v(x)$.\par
The bifurcation of the fractal SPT is given by the coarse-graining of the polynomial $w(x)$ (Eq. \ref{polycoarsegrain}). Because this is coarse-graining, the degree of $w(x)$ decreases every ERG step. Eventually, we are left with only three possibilities: $w(x)=1$, $x$, or $\bar{x}$. The first corresponds to $H_A$, and the latter two are $H_\pm$.\par
Next, we return to the $y$ shift in one of the sublattices for general fractal SPTs. For an ERG step with any $l$, we can shift by $-l$, perform the step as above for $l=0$, and shift back. Because we are now on the coarse-grained unit cell, however, the $l$'s on our bifurcated models are smaller than before. Again, this means that we only need to consider $l=1$,$-1$, and $0$. The $l=0$ case has already been done. The $l=\pm 1$ cases introduce new potential fixed points, which we label $H^{\pm 1}_{(A,+,-)}$, where the superscript indicates the value of $l$. We find, however that these are not fixed points:
\begin{align}
H^{\pm 1}_{A}(a) \simeq H_A(2a)+H_A(2a),\\
H^{\pm 1}_{\pm}(a)=H_\pm(2a)+H_A(2a).
\end{align}
In other words, $l$ being $\pm 1$ does not affect the ERG. The local gates and twist phases are not the same as the $l=0$ case, however. In particular, when $l=\pm 1$, the bifurcated copies are placed on different qubits within the unit cell than in the $l=0$ ERG. Alternatively, this can be thought of as locally relabelling the lattice, which (unlike global sublattice shifts) is allowed in the ERG. The details of this process are in Section 2 of the Mathematica notebook.\par 

It is relatively simple to describe the local gates required to reach these fixed points. As mentioned in Section \ref{sec:fractal}, the sets $Q_A$ and $Q_B$ are given by ${Q_A(x,y) = c(x,y)P_A(x,y)}$ and ${Q_B(x,y) = d(x,y)P_B(x,y)}$, with $c$ and $d$ arbitrary polynomials, and
\begin{align}
    P_A(x,y)= 1 + \bar{y}f(\bar{x}),\quad P_B(x,y)=1+yf(x).
\end{align}
In Section 4 of the Mathematica notebook, we show how to arrange these gates to bifurcate $H_A$.

What about general fractals? These may be deduced from the bifurcation of $H_A$, as follows. First, if $w(x)$ is a monomial, then for any $l$, Eq. \ref{higherfractal} is just $H_A$, but with one of the sublattices shifted by some $y^lx^k$.\par
Now, we have a set of circuits each labelled by $(Q_A,Q_B)$, which split $H_A$ into two copies of itself. If instead, we take $(Q_A,y^lx^kQ_B)$, our monomial model with $w(x)=x^k$ then bifurcates into models of the form
\begin{align}\label{monomialerg}
    H=\begin{pmatrix}
    0 & 1 \\
    1 & 0 \\
    y^{l'} x^{k'}(1+yf(x)) & 0 \\
    0 & \bar{y}^{l'} \bar{x}^{k'}(1+\bar{y}f(\bar{x}))
    \end{pmatrix},
\end{align}
where the parameters $l'$ and $k'$ correspond to shifts of one sublattice by $l$ and $k$, but in the coarse-grained coordinates. In other words, a monomial $w(x)$ leads to bifurcation into two other monomials.

We now extend this to general polynomials $w(x)$. First, notice that the bipartite graph corresponding to such a model can be made by taking the edges of a set of monomial models and adding them mod 2. This set of monomial models is given by the terms in $w(x)$. Next, consider the set of symmetric circuits given by $(Q_A,y^lw(x)Q_B)$. These polynomials can also be made by adding individual polynomials of the form $(Q_A,y^lx^kQ_B)$ mod 2, which is equivalent to multiplying the circuits since $CZ=CZ^{-1}$. We then see that doing the ERG on the polynomial model with gates given by $(Q_A,y^lw(x)Q_B)$, is the same as doing the ERG on each monomial model with gates $(Q_A,y^lx^kQ_B)$, and then adding the cluster state graphs mod 2. Therefore, the flow of a cluster FSPT with general polynomial $w(x)$ is simply determined by the flow for each of the monomial terms in $w(x)$. 
The result is the bifurcation given in Eq.~\eqref{polycoarsegrain}. 

We have shown that there are only three fixed points. It is obvious from Eq. \ref{polycoarsegrain} and Eq. \ref{fractaltwistphasecg} that $H_A$ is self-bifurcating, but it is not clear that $H_\pm$ cannot self-bifurcate. We now show that this cannot happen for $H_+$ ($H_-$ is similar). This argument is also similar to the one we used to show the linear cluster state does not self-bifurcate.

\begin{figure}[t!]
    \centering
    \includegraphics[width=\columnwidth]{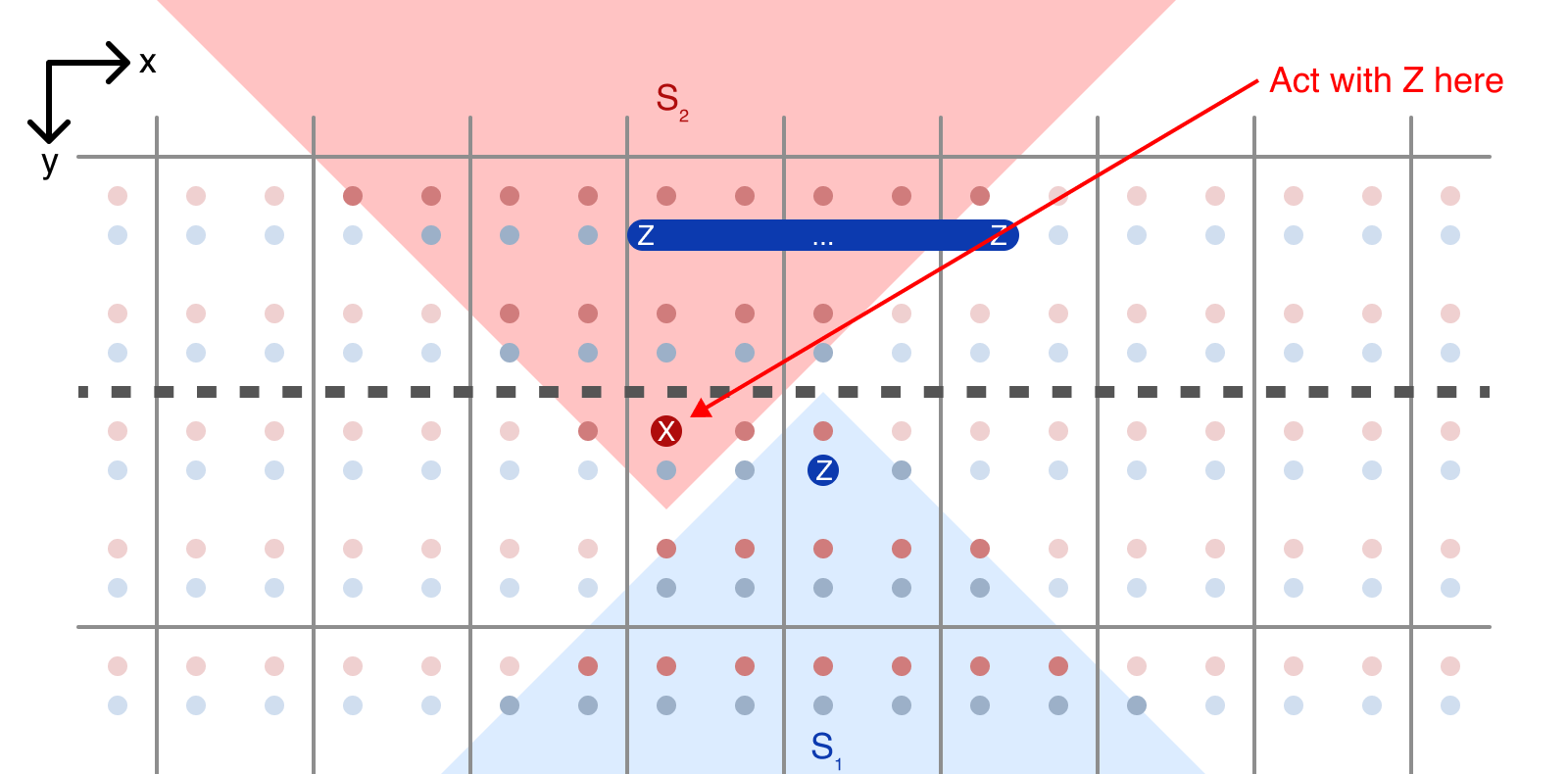}
    \caption{Calculation of twist phases for one copy of $H_+$. The truncated symmetry $S_1$ anticommutes with only one local term (represented schematically for an arbitrary polynomial $f(x)$). An equivalent action on $H_+(2a)$ is a single Pauli $Z$ gate. This anticommutes with one symmetry, $S_2$. For the same commutation relation, we place a $Z$ operator on the row where the symmetry only acts on one qubit. On the coarse-grained lattice (grid lines), the $Z$ operator is always on a different lattice site.}
    \label{fractaltwistphasefig}
\end{figure}

First, from Eq. \ref{fractaltwistphasecg}, the twist phases of $H_+$ are
\begin{align}
    W^{(y)}=\begin{pmatrix}
    0 & 0 & 0 & 1 \\
    0 & 0 & x & 0 \\
    0 & \bar{x} & 0 & 0 \\
    1 & 0 & 0 & 0
    \end{pmatrix}.
\end{align}
The important detail is that $W^{(y)}_{14}=1$, which means that the first and fourth symmetries, $S^{(1)}$ and $S^{(4)}$, only anticommute if they are on the same lattice site (using the basis we defined previously by coarse-graining Eq. \ref{fractalsymspoly}). However, this is impossible with only copies of $H_+$. To see this, we cut along $y=0$. All of the symmetries act on this line at only a single point. Because of this, they only anti-commute with a single local term of any copy of $H_+$ (Fig. \ref{fractaltwistphasefig}). This means we can write the action along the cut as a single Pauli $Z$ operator acting on an adjacent coarse-grained lattice site. Therefore, symmetries cannot have nontrivial twist phases with symmetries acting on the same site. If we allow for a copy of $H_A$, however, this problem is resolved, as the local terms for $H_A$ allow for Pauli $Z$'s on the same lattice site.

\end{document}